\documentclass[aip,cha,amsmath,amssymb,10pt,twocolumn,numerical,reprint]{revtex4-1}
\usepackage{psfrag}
\usepackage[dvips]{graphicx}
\usepackage[english]{varioref}
\usepackage[english]{babel}
\usepackage{amsmath}
\usepackage{amsfonts}
\usepackage{eucal}
\usepackage{amssymb,amsmath}
\usepackage{natbib}

\newcommand{\mh}{\mathbf}

\newcommand{\tn}{\textnormal}
\newcommand{\bos}{\boldsymbol}
\newcommand{\defeq}{\mathrel{\mathop:}=}
\newcommand{\eqdef}{=\mathrel{\mathop:}}
\newcommand{\mb}{\mathbb}
\newcommand{\mc}{\mathcal}
\newcommand{\defeql}{\ \mathrel{\mathop:}=\ }
\newcommand{\eqdefl}{\ =\mathrel{\mathop:}\ }
\newcommand{\eq}{\ =\ }
\newcommand{\qc}{\quad,}
\newcommand{\qsc}{\quad;}
\newcommand{\qd}{\quad.}
\newcommand{\qcq}{\quad,\quad}

\begin{document}

\title{Oseledets' Splitting of Standard-like Maps}

\author{M. Sala}\email[]{msala@pks.mpg.de}
\affiliation{Max Planck Institute for the Physics of Complex Systems,\\
N\"{o}thnizer Stra{\ss}e 38, 01187 Dresden, Germany}
\author{R. Artuso}\email[]{roberto.artuso@uninsubria.it}
\affiliation{Center for Nonlinear and Complex Systems and Dipartimento di Scienza ed Alta Tecnologia,\\Universit\`a degli Studi dell'Insubria, Via Valleggio 11, 22100, Como, Italy}
\affiliation{Istituto Nazionale di Fisica Nucleare, Sezione di Milano,\\Via Celoria 16, 20133 Milano, Italy}


\begin{abstract}
For the class of differentiable maps of the plane and, in particular, for standard-like maps (McMillan form), a simple relation is shown between the directions of the local invariant manifolds of a generic point and its contribution to the finite-time Lyapunov exponents (FTLE) of the associated orbit. By computing also the point-wise curvature of the manifolds, we produce a comparative study between local Lyapunov exponent, manifold's curvature and splitting angle between stable/unstable manifolds. Interestingly, the analysis of the Chirikov-Taylor standard map suggests that the positive contributions to the FTLE average mostly come from points of the orbit where the structure of the manifolds is \emph{locally hyperbolic}: where the manifolds are flat and transversal, the one-step exponent is \emph{predominantly} positive and large; this behaviour is intended in a purely statistical sense, since it exhibits large deviations. Such phenomenon can be understood by analytic arguments which, as a by-product, also suggest an explicit way to point-wise approximate the splitting.
\end{abstract}

\maketitle

\begin{quotation}
The invariant manifolds embody the structure of chaotic dynamical systems, whose non-linear strength is quantified by the Lyapunov exponents.\\ Alongside the magnitude of the exponents, the structural features of such systems are encoded in the non-uniformity of their un/stable manifolds: \emph{how frequently} the two are transversal to each other and \emph{how much} their shapes are curvilinear can be regarded as their two simpler descriptors.\\By investigating Lyapunov exponents, splitting angles and manifold's curvatures for a 2D non-hyperbolic map, we find here strong correlations between the three observables which allow to quantify and characterise the system deviations from a uniformly hyperbolic behaviour.
\end{quotation}

\section{Introduction}
In the huge literature concerning the study of 2D maps much effort has been devoted to the study of \emph{Lyapunov exponents} and \emph{invariant manifolds}, the former being the primary indicator for chaos, the latter representing the true skeleton of phase-space. Lyapunov exponents are the probes that measure the stretch \& fold mechanism and the associated loss of information, quantified by the KS-entropy; on the other hand, invariant manifolds of \emph{fixed-points} represent the underlying geometry and are essential in the construction of invariant measures and generating partitions for symbolic coding the dynamics.\\
In this paper we focus on the phase-space structure at \emph{generic} points: there, one can associate two left-invariant curves (not invariant, in general) whose linearisations are called covariant Lyapunov vectors (CLV) \cite{Wolfe,Ginelli} and form the so-called {Oseledets' splitting} \cite{osel,ruel}. In 2D systems it is still computationally reasonable to consider also the point-wise {curvature}, since this is essentially a single scalar quantity for each of the curves and no new information comes from their normal vectors (indeed, these simply compose the \emph{dual} basis of the Lyapunov vector basis).\\
On this grounds, a \emph{local} characterization of a generic phase-space point would start by including three types of observables: the \emph{one-step Lyapunov exponents} (whose time-average along an arc of orbit produces the \emph{finite-time Lyapunov exponents} (FTLE)), the \emph{splitting angle} between stable/unstable manifolds and the \emph{curvatures} of the manifolds at that same point. By the well known symmetries of standard-like maps which connect stable and unstable manifolds, here we can limit ourselves to a minimal set of three observables: a one-step exponent and a curvature for a single (say, unstable) manifold, plus the splitting angle. Through the collection of their phase-space distributions and producing global statistics, we are then able to characterise the relations between the \emph{large deviations} of the FTLE \cite{Tom,Ces,L1,L2,Mel}, the presence of \emph{very small} splitting angles and the \emph{flatness} of the manifolds; this analysis allows to formulate non-trivial considerations about the structure of a system. Once applied to the Chirikov-Taylor map (CT), the emerging picture turns out to be quite intuitive: the values of the one-step exponent that contribute more to the FTLE sum mainly occur at points where the manifolds are almost orthogonal and the manifolds are essentially flat lines.
The paper is organized as follows: in section \ref{slmap} we introduce the basic properties of standard-like maps and their stability structure; in section \ref{geostab0} we illustrate the relations between geometry and stability of generic orbits, exposing the employed algorithms. Finally, section \ref{numex} is devoted to the numerical results on the CT map, along with their explanations through some analytic results.

\newpage
\section{\label{slmap}Standard-like maps}
We consider symplectic \emph{standard-like} maps of the plane $\mh{\Phi}:\mb{R}^2\to\mb{R}^2$ in the form introduced by McMillan in order to model generic \emph{beam-focusing} systems\cite{mcmill,foot}:
\begin{align}
\mh{\Phi}:
\left(\begin{array}{c}
x\\
y
\end{array}\right)
\mapsto
\left(\begin{array}{c}
f(x)-y\\
x
\end{array}\right)\qc
\label{stan}
\end{align}
with the function $f\in\mathcal{C}^2(\mathbb{R})$ representing the net action of a sequence of nonlinear lenses in a periodic cavity. For any choice of $f$, map (\ref{stan}) is \emph{area preserving} and \emph{reversible}, namely there exists an involution $\mh{S}$ (a map whose square is the identity, $\mh{S}^2=id$) that \emph{conjugates} $\mh{\Phi}$ to its inverse:
\begin{align}
\mh{S}\circ\mh{\Phi}\ =\ \mh{\Phi}^{-1}\circ\mh{S}\quad.
\label{inv}
\end{align}
Our choice of coordinates is motivated by the simple form taken by the associated involution: $\mh{S}(x,y)^T=(y,x)^T\,$, represented by matrix $\mh{S}=\left[\begin{smallmatrix}0&1\\1&0\end{smallmatrix}\right]$ and corresponding to the {reflection} about the diagonal line. This implies that all the fixed-points of map (\ref{stan}) lie along such line, $x_{\tn{fix}}=y_{\tn{fix}}$, while all the periodic points $\mh{x}_*$ share the same period with their reflections $\mh{S}\mh{x}_*$. In this sense, the pair $(x,y)$ is a more convenient representation for the geometry of the usual standard mapping involving the canonical pair $(x,p)\defeq(x,x-y)$ :
\begin{align}
\left(\begin{array}{c}
x\\
p
\end{array}\right)
\mapsto
\left(\begin{array}{c}
x+p+F(x)\\
p+F(x)
\end{array}\right)\qc
\label{sta}
\end{align}
where the function $F(x)\defeq f(x)-2x$ is then physically interpreted \cite{Chirikov} as an external \emph{force} acting on a point mass which, in case of integrable $f$, is ruled by a Hamilton function with potential $V(x)=x^2-\int\!f\tn{d}x$ which is \emph{delta-kicked} in time. In such setting, fixed-points correspond to \emph{equilibria}, where both the force and momentum vanish:
\begin{align}
F(x_{\tn{fix}})=0\qcq y_{\tn{fix}}=x_{\tn{fix}}\qd
\label{fix}
\end{align}
Notice that the two representations are equivalent only when defined on the same \emph{geometry}; instead, in case of \emph{periodic} position $x\in\mb{S}$ and \emph{unbounded} momentum $p\in\mb{R}$, the correspondence is {broken}: the pair $(x,y)\in\mb{T}^2$ belongs to a 2-torus while the pair $(x,p)\in\mb{S}\times\mb{R}$ belongs to a cylinder. In such case the isomorphism between the two representations would require the additional definition of a \emph{winding number} $N\in\mb{Z}$ in order to track the growth of momentum: $p=x-y+2\pi N$ (assuming a period of $2\pi$).
As a numerical benchmark for mixed phase-space, we consider here the Chirikov-Taylor standard map (CT) \cite{Chirikov,lieb} on the 2-torus, but we avoid to address the associated diffusion of momenta; such map is defined as:
\begin{align}
\mh{\Phi}:
\left(\begin{array}{c}
x\\
y
\end{array}\right)
\mapsto
\left(\begin{array}{c}
2x\ +\ K\sin(x)\ -\ y\\
x
\end{array}\right)_{\tn{mod}\,2\pi}\qc
\label{CT}
\end{align}
by the choice $f(x)\eq2x\ +\ K\sin(x)$ inside of map (\ref{stan}). To characterize any tangent structure effectively, indeed, it is more convenient to consider \emph{bounded} orbits: these may be closed (\textit{e.g.} quasi-periodic cycles) or open orbits on a bounded phase-space (\textit{e.g.} chaotic orbits on a torus); in the case of unbounded orbits, the tangent information can still be localized in phase-space, but becomes more difficult to track it down for direct observations. Along with the symmetry properties illustrated above, this is also why we employ the pair $(x,y)\in\mb{T}^2$, avoiding the computation of the winding number. With due care, the very same approach can be applied also to the canonical $(x,p)$ representation on the 2-torus, as discussed in \cite{Proc}.

\subsection{Stability}
Given the structure of map (\ref{stan}), the associated Jacobian matrix depends only on the \emph{first} coordinate:
\begin{align}
\mh{J}(\mh{x})\ \defeq\ \left.\frac{\partial\mh{\Phi}}{\partial\mh{x}}\right|_{\mh{x}}=\left[\begin{array}{cr}f'(x)&-1\\1&0\end{array}\right]\ \equiv\ \mh{J}(x)\qc
\label{jac}
\end{align}
with the prime symbol standing for the $x$ derivative. As a consequence, also $\mh{J}$ represents a reversible (now \emph{linear}) transformation between tangent spaces:
\begin{align}
\left(\mh{J}\right)^{-1}\ =\ \mh{S}\,\mh{J}\,\mh{S}\qsc
\label{jacinv}
\end{align}
indeed, $\mh{J}$ has determinant $1$ as the map $\mh{\Phi}$ is area-preserving. By defining $\mh{x}_n\defeq\mh{\Phi}^n(\mh{x})$ and $\mh{J}_n\defeq\mh{J}(\mh{x}_n)$, the $n$-th iterate Jacobian matrix is written as a product:
\begin{align}
\left.\frac{\partial\mh{\Phi}^{n}}{\partial\mh{x}}\right|_\mh{x}\ \eqdef\ \mh{F}^{n}(\mh{x})\ =\ \mh{J}_{n-1}\,\dots\,\mh{J}_0\qc
\label{jacF}
\end{align}
revealing its \emph{exponential} dependence on the orbit length $n$ and inducing the fundamental \emph{cocycle} property:
\begin{align}
\mh{F}^{n+m}\eq\mh{F}^{n}\circ\mh{\Phi}^m\ \mh{F}^{m}\qd
\label{coc}
\end{align}
Trivially then, also each matrix $\mh{F}^n$ has determinant $1$. Since property (\ref{inv}) is inherited by all the map iterates:
\begin{align}
\mh{S}\circ\mh{\Phi}^{n}\ =\ \mh{\Phi}^{-n}\circ\mh{S}\qc
\label{inv2}
\end{align}
relation (\ref{jacinv}) can be generalized to any $n$, so that:
\begin{align}
\mh{F}^{n}(\mh{x})\ =\ \mh{S}\,\mh{F}^{(-n)}(\mh{S}\mh{x})\,\mh{S}\qc
\label{jac2}
\end{align}
the two Jacobian matrices of the iterated map $\mh{\Phi}^n$ from point $\mh{x}$ and of its inverse $\mh{\Phi}^{-n}$ from the reflected point $\mh{S}\mh{x}$ are {connected} by the linear transformation $\mh{S}$.

\subsubsection{Fixed Points}
Since map (\ref{stan}) is area preserving, the stability of its {fixed points} $\mh{x}_*$ depends only on the trace of matrix $\mh{J}$, $f'_*\equiv\tn{tr}(\mh{J}(x_*))=f'(x_*)$; this gives the following form for its eigenvalues $\chi_\pm$ :
\begin{align}
\chi_\pm\ =\ \tn{sign}(f'_*)\cdot
\left\{\begin{array}{l}
	e^{\pm\tn{arccosh}|f'_*/2|}\qcq|f'_*/2|\geq1\\
	e^{\pm i\,\tn{arccos}|f'_*/2|}\qcq|f'_*/2|\leq1
\end{array}\right.
\label{stab}
\end{align}
and its eigenvectors $\mh{w}_\pm$ :
\begin{align}
\mh{J}\,\mh{w}_\pm=\mh{w}_\pm\,\chi_\pm\ \Rightarrow\ \mh{w}_\pm
\propto
\left[
\begin{array}{c}
\chi_\pm\\
1
\end{array}
\right]
\propto
\left[
\begin{array}{c}
\cos(\alpha_\pm)\\
\sin(\alpha_\pm)
\end{array}
\right]\ ;
\label{stab2}
\end{align}
In the unstable case $|f'_*/2|>1$ , the eigenvalues are real and correspond to $\chi_\pm=\cot(\alpha_\pm)$, with $\alpha_\pm$ the angles of slope of the eigenvectors. Then, the symmetry of the eigenvalues $\chi_-=1/\chi_+$ implies that $\alpha_-=\frac{\pi}{2}-\alpha_+$, that is, the eigenvectors at the fixed point  $\mh{x}_*$ (which is on the diagonal) are \emph{symmetric} about the diagonal.\\
In general, the angle between the two vectors is called \emph{splitting angle}, $\theta=(\alpha_--\alpha_+)$, which, in the unstable case, can be written as $\theta=\frac{\pi}{2}-2\alpha_+$; in these terms, the parabolic case $|f'_*/2|=1=|\chi_\pm|$ corresponds to the \emph{degenerate} limit $\alpha_\pm=\frac{\pi}{4}\ \Rightarrow\ \theta=0$, where both the eigenvectors {align} to the diagonal and the splitting angle vanishes. The stable case $|f'_*/2|<1$ implies complex eigenvectors, whose splitting angle cannot be defined.\\
By standard results \cite{Pesin}, we know that the eigenvectors at an unstable fixed-point correspond to the \emph{linearisations} of the associated \emph{invariant manifolds}; then, for the $(x,y)$ choice of representation, the manifold's slope angles $\alpha_\pm$ at $\mh{x}_*$ are directly connected to the \emph{strength} of instability through the definition of Lyapunov exponents (LE):
\begin{align}
\lambda_\pm\ \defeq\ \ln|\chi_\pm|\eq\ln|\cot(\alpha_\pm)|\qd
\label{stab3}
\end{align}
Notice that, according to (\ref{stab3}), when the manifolds at some fixed-point are horizontal/vertical (\textit{i.e.} $\alpha_\pm=0/\frac{\pi}{2}$) the associated Lyapunov exponents diverge to $\pm\infty$. In section \ref{geostab0} we show how, in the $(x,y)$ coordinates, such a non-trivial interplay between geometry and stability holds exactly also for \emph{generic} points, in both dynamical cases of chaotic and quasi-periodic orbits. Interestingly, relation (\ref{stab3}) can be extended to generic 2D maps and arbitrary coordinates (see Appendix \ref{appx:a1}).

\subsection{Left-invariant Curves}
Here we focus on the \emph{local} {generalisation} of the concept of {invariant manifolds} extended to arbitrary \emph{non-fixed} points \cite{Pesin}; this leads to the definition of phase-space sub-sets which are \emph{not} exactly invariant:
\begin{align}
\mathcal{P}^\pm(\mh{x})=\left\{\mh{y}:\lim_{n\rightarrow\mp\infty}\left\| \mh{\Phi}^n(\mh{x})-\mh{\Phi}^n(\mh{y})\right\|=0\right\},
\label{set}
\end{align}
but fulfil the so-called \emph{left-invariant} relation:
\begin{align}
	\mh{\Phi}^n\left(\mathcal{P}^\pm(\mh{x})\right)\eq\mathcal{P}^\pm \left(\mh{\Phi}^n(\mh{x})\right)\qd
	\label{cov}
\end{align}
Such sets are \emph{transported} along with the dynamics of $\mh{x}$ and, in the limits $n\to\pm\infty$, \emph{converge} to its orbit; for this reason, these are called local- or left-invariant manifolds.\\
By assuming smooth maps $\mh{\Phi}$, it is possible to show \cite{Pesin} that, \emph{locally}, the sets $\mathcal{P}^\pm(\mh{x})$ can be represented by two differentiable curves $\bos{\gamma}(s)$ parametrized by their \emph{arc length} $s$ and such that $\bos{\gamma}(0)=\mh{x}$. With respect to the invariant measure (here, Lebesgue) these left-invariant curves are then \emph{almost-everywhere} smooth, since they may exhibit arbitrary numbers of \emph{isolated singularities} \cite{Sinai}; operatively, this does not pose serious obstructions to the \emph{numeric observation} of such curves, since the probability to fall \emph{exactly} upon singularities (\textit{e.g.} cusps) is zero.\\This is the guiding idea of our approach: the invariant structure of a non-uniform chaotic system may have a \emph{dense} set of pathologies but, as long as these have zero-measure, its phase-space dependence can still be probed.

\subsubsection{Curves Stability}
An important consequence of reversibility (\ref{inv}) applied to (\ref{set}) is that curves of opposite stability are connected:
\begin{align}
	\mh{S}\mathcal{P}^\pm\left(\mh{S}\mh{x}\right)\ =\ \mathcal{P}^\mp(\mh{x})\quad,
	\label{revcov}
\end{align}
so that the set $\mathcal{P}^+$ associated to point $\mh{x}$ corresponds to the reflection of the set $\mathcal{P}^-$ for the reflected point $\mh{S}\mh{x}$. Parallel to equation (\ref{jac2}), this fact generalizes the results in \cite{dev} about fixed points and allows to deduce properties of sets $\mathcal{P}^-$ from those of sets $\mathcal{P}^+$. On such grounds, let us consider a single type of set $\mathcal{P}(\mh{x}_0)$ (say, $\mathcal{P}^+$), and represent it by a parametric curve $\bos{\gamma}_0(s)$ with arc length $s$ and base-point $\bos{\gamma}_0(0)=\mh{x}_0$. Then, by relation (\ref{cov}), the sequence of curves:
\begin{align}
\{\,\bos{\gamma}_n(s)\,\}_{n=0..\infty}\qcq\bos{\gamma}_n(0)=\mh{\Phi}^n(\mh{x}_0)\qc
\label{curves}
\end{align}
coincides with the sequence of sets $\left\{\mathcal{P}\left(\mh{\Phi}^{n}(\mh{x}_0)\right)\right\}$ so that, for some initial $s_{[0]}$, the image $\mh{\Phi}\left(\bos{\gamma}_0\left(s_{[0]}\right)\right)$ lies in $\mc{P}(\mh{x}_1)$ and so there exists a \emph{new} arc length $s_{[1]}$ for which:
\begin{align}
\bos{\gamma}_1\left(s_{[1]}\right)\eq\mh{\Phi}\left(\bos{\gamma}_0\left(s_{[0]}\right)\right)\qd
	\label{revcov2}
\end{align}
This leads to define a \emph{sequence of mappings} $\varphi_{[n]}:\mathbb{R}\rightarrow\mathbb{R}$ such that $s_{[n+1]}=\varphi_{[n]}\left(s_{[n]}\right)$ and relation (\ref{revcov2}) becomes:
\begin{align}
\bos{\gamma}_{n+1}\left(\varphi_{[n]}\left(s_{[n]}\right)\right)\ =\ \mh{\Phi}\left(\bos{\gamma}_{n}\left(s_{[n]}\right)\right)\qsc
\label{curve}
\end{align}
each mapping $\varphi$ is said to be \emph{semi-conjugated} to map $\mh{\Phi}$. Notice that, in the case of $\mh{x}_0\equiv\mh{x}_*$ \emph{periodic}, $\mh{\Phi}^P(\mh{x}_*)=\mh{x}_*$ for some $P\in\mb{N}$, all the curves collapse over a single one, precisely one of the two \emph{exact} invariant manifolds of $\mh{x}_*$. By denoting the derivative with respect to $s_{[n]}$ by an upper dot $\dot{\ }$, we have $\|\dot{\mh{x}}_n\|=1\,,\,\forall\,n$ since {each} $s_{[n]}$ is an arc length, and deriving (\ref{curve}) produces a map between unit tangent vectors:
\begin{align}
\dot{\bos{\gamma}}_{n+1}\left(s_{[n+1]}\right)\ \dot{\varphi}_{[n]}\left(s_{[n]}\right)\ =\ \mh{J}_n\ \dot{\bos{\gamma}}_{n}\left(s_{[n]}\right)\quad,
\label{curvetan}
\end{align}
which are tangent to two consecutive left-invariant sets; notice that last equation follows from the chain-rule of derivatives applied to equation (\ref{curve}), so the matrix $\mh{J}_n$ is the Jacobian of map $\mh{\Phi}$ evaluated at point $\bos{\gamma}_{n}(s_{[n]})$. Rewriting (\ref{curvetan}) more compactly:
\begin{align}
\dot{\bos{\gamma}}_{n+1}\,\dot{\varphi}_{[n]}=\mh{J}_n\,\dot{\bos{\gamma}}_{n}\qc
\label{curvetan2}
\end{align}
we deduce that the derivative of the arc length mapping:
\begin{align}
\dot{\varphi}_{[n]}(s_{[n]})\ =\ \frac{\tn{d}s_{[n+1]}}{\tn{d}s_{[n]}}\qc
\end{align}
represents the \emph{local} expanding factor along the curve $\mathcal{P}(\mh{\Phi}^{n}(\mh{x}_0))$. This means that, by definition (\ref{jacF}), repeated iteration of (\ref{curvetan2}) for $n$ times starting from $\bos{\gamma}_0$ yields:
\begin{align}
\dot{\bos{\gamma}}_n\left(\,\prod_{q=0}^{n-1}\dot{\varphi}_{[q]}\,\right)\eq\mh{F}^{n}(\mh{x}_0)\,\dot{\bos{\gamma}}_0\qc
\label{ktang}
\end{align}
and provides a way to re-write the \emph{finite-time Lyapunov exponent} (FTLE) for a unit tangent vector $\mh{w}_0\equiv\dot{\bos{\gamma}}_0$ :
\begin{align}
\chi^{n}(\mh{x}_0,\mh{w}_0)\defeql\tfrac{1}{n}\ln\left\|\mh{F}_{n}(\mh{x}_0)\mh{w}_0\right\|\qc
	\label{ftle}
\end{align}
as the Birkhoff average of the local expanding factors:
\begin{align}
\chi^{n}(\mh{x}_0,\dot{\bos{\gamma}}_0)\eqdefl\chi^n_0\eq\tfrac{1}{n}\sum_{q=0}^{n-1}\ln\left|\dot{\varphi}_{[q]}\right|\qc
	\label{ftle4}
\end{align}
in analogy with the FTLE expression for 1D maps \cite{lieb}.\\
Notice that, for 2D systems, the FTLE (\ref{ftle}) is computed by taking a random unit vector $\mh{w}_0$, since $\dot{\bos{\gamma}}_0$ is unknown: in the limits $n\to+/-\infty$ one gets respectively the largest/smallest Lyapunov exponent (for area-preserving maps, these are equal in modulus and opposite in sign).

\subsubsection{Covariant Lyapunov Vectors}
The \emph{covariant Lyapunov vectors} (CLV) are defined \citep{Ginelli} to be the \emph{unique} left-invariant (and phase-space dependent) vectors producing the \emph{same} Lyapunov exponent (\ref{ftle}) in \emph{both} of the two temporal limits: this implies that the CLV associated to positive/negative exponent is the precise tangent direction which respectively converges to zero in the limit $k\rightarrow\mp\infty$. By the Oseledets' theorem \cite{osel}, these form a tangent basis \emph{almost-everywhere} in phase-space, the so-called \emph{Oseledets' splitting} of tangent space; then, the comparison of the CLV definition with (\ref{set}) and (\ref{ktang}) leads to identify the linearisations $\dot{\bos{\gamma}}^\pm$ of left-invariant curves as the Oseledets' splitting itself \cite{osel,ruel}. Therefore, the ideas introduced hereafter are equivalent alternatives and, possibly, extensions of the existing results based on the Lyapunov vectors analysis \cite{Benettin,Wolfe,Ginelli}.

\section{\label{geostab0}Geometric Stability}
\begin{figure}[t!]
\includegraphics[width=.48\textwidth]{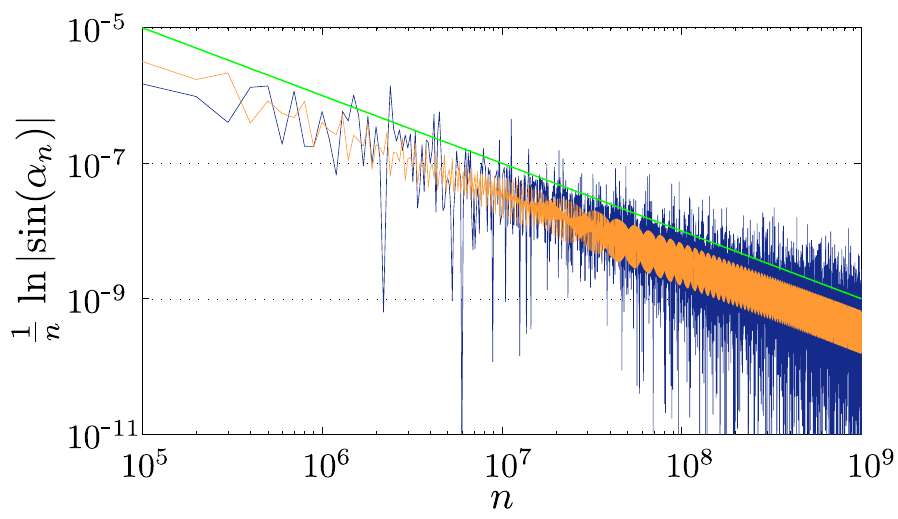}
\caption{\label{diffang1}Decay of term (\ref{ftleterm}) from relation (\ref{ftle2}), which allows to reduce the FTLE computation from (\ref{ftle2}) to (\ref{ftleOK}). The data come from $10^9$ iterations (sampled each $10^5$) of the CT map (\ref{CT}) with parameter $K=\pi/4$ (very weak chaos); the orange/blue graph is respectively for a regular/chaotic orbit (cases (2)/(4) in figure \ref{comp1}), the reference green line is $1/n$.}
\end{figure}

The method employed here, already introduced in \cite{Giona} for different purposes, exploits the isomorphism between special linear transformations (the group $SL(2,\mathbb{R})$) and rational functions (the \emph{M\"obius} group, see appendix \ref{appx:a1}). Unit tangent vectors $\dot{\bos{\gamma}}$ can be uniquely represented by polar angles $\alpha\in[-\pi,\pi]$ through the two coordinates:
\begin{align}
\psi\ \defeq\ \cot(\alpha)\qcq\sigma\ \defeq\ \rm{sign}(\alpha)\qc
\label{coo}
\end{align}
which allow to cover the whole unit circle:
\begin{align}
\dot{\bos{\gamma}}
\ \equiv\ 
\left[
\begin{array}{c}
	\dot{x}\\
	\dot{y}
\end{array}
\right]
\eq
\left[
\begin{array}{c}
	\cos(\alpha)\\
	\sin(\alpha)
\end{array}
\right]
\eq
\frac{\sigma}{\sqrt{1+\psi^2}}\left[
\begin{array}{c}
	\psi\\
	1
\end{array}
\right]\qd
	\label{psi}
\end{align}
These are functions of the arc length $s$, as vector $\dot{\bos{\gamma}}$ is, and thus functions of phase-space (arc lengths $s$ and points $\mh{x}$ are related by the choice of a left-invariant curve $\bos{\gamma}$); from here on, the shorthand $\psi(\mh{x}_n)\equiv\psi_n$ is used, also for any function with the same type of dependence.\\
By inserting the explicit form of Jacobian matrix (\ref{jac}) into the tangent evolution (\ref{curvetan}) and making use of (\ref{psi}), the corresponding evolutions for $\psi$ and $\sigma$ are easily deduced:
\begin{align}
\psi_{n+1}\ &=\ f'(x_n) - \frac{1}{\psi_n}\ \ ,\label{psievo}\\
\sigma_{n+1}\ &=\ \rm{sgn}(\psi_n)\,\sigma_n\qd
\label{sigevo}
\end{align}
While (\ref{sigevo}) is essentially a logic relation, evolution (\ref{psievo}) encodes the \emph{whole} linearised dynamics; {whenever} the latter is \emph{expansive}, we can show that it \emph{converges} from any random $\psi_0$ to the \emph{true} left-invariant slope $\psi_n\equiv\psi(\mh{x}_n)$ (see section \ref{numcon}). Before going into convergence issues, we notice that the representation (\ref{psi}) allows to write the local expanding factors explicitly in terms of the slopes:
\begin{align}
\dot{\varphi}_{[n]}\eq\frac{\dot{x}_n}{\dot{y}_{n+1}}\eq\psi_n\frac{\dot{y}_n}{\dot{y}_{n+1}}\eq|\psi_n|\sqrt{\frac{1+\psi_{n+1}^2}{1+\psi_{n}^2}}\qsc
	\label{phi}
\end{align}
the modulus on $\psi_n$ in last equality (due to (\ref{psi}) and (\ref{sigevo})) means that $\dot{\varphi}$ never changes sign, since $\varphi$ is conjugated to $\mh{\Phi}$ and thus invertible and monotone. It is then remarkable how the \emph{geometric} quantities $\psi_n$ associated to the local manifolds also determine the local \emph{stability}; indeed, by equation (\ref{phi}) we can re-write the FTLE in (\ref{ftle}) as:
\begin{align}
\chi^n_0\eq\tfrac{1}{n}\sum_{q=0}^{n-1}\ln\left|\psi_q\right|\ -\ \tfrac{1}{n}\ln\left|\frac{\sin(\alpha_n)}{\sin(\alpha_0)}\right|\qc
\label{ftle2}
\end{align}
by inserting the second line of (\ref{psi}), $\dot{y}=\sin(\alpha)$, into (\ref{phi}).\\
At this stage, we observe (also numerically, see figure \ref{diffang1}) that the last term in (\ref{ftle2}) goes to zero in the $n\rightarrow\infty$ limit and can thus be discarded from the FTLE: notice first that it equally depends upon both the angles $\alpha_0/\alpha_n$ at the initial/final points of the orbit. As a consequence, it always contains a spurious contribution due to the random choice $\psi_0\equiv\cot(\alpha_0)$, which corresponds to the unavoidable choice of an initial random vector for \emph{any} tangent evolution to compute the FTLE \cite{lieb}. This may already give a clear motivation to consider:
\begin{align}
-\ \tfrac{1}{n}\ln\left|\frac{\sin(\alpha_n)}{\sin(\alpha_0)}\right|\qc
\label{ftleterm}
\end{align}
as the typical \emph{transient} data which drops as $\tfrac{1}{n}$ in any orbit-wise calculation of the FTLE.
To clarify this point, notice further that the only cases in which the infinite-time limit of term (\ref{ftleterm}) may exhibit problems arise when the function $\sin(\alpha_n)$ vanishes, \textit{i.e.} when $\alpha_n\to 0,\pi$ and the local manifold becomes horizontal as $n\to\infty$; but, as already pointed out, the case in which the local manifold is \emph{asymptotically} horizontal is impossible, because it would imply that the Lyapunov exponent itself diverges:
\begin{align}
\alpha_n\rightarrow0\quad\Rightarrow\quad\ln\left|\psi_n\right|\rightarrow\infty\quad\Rightarrow\quad\chi^n_0\rightarrow\infty\qsc
\label{ftle5}
\end{align}
and this is ruled out by the invertibility of map $\mh{f}$. As an ultimate check (not shown) we have also calculated the FTLE by both standard techniques and expression (\ref{ftle2}) and verified that the magnitude of term (\ref{ftleterm}) (exemplified in figure \ref{diffang1}) is exactly the same of the FTLE fluctuations due to different initial perturbations, or, $\alpha_0$.

\subsection{\label{geostab} Slopes \& Exponents }
\begin{figure}[h!]
\includegraphics[width=.48\textwidth]{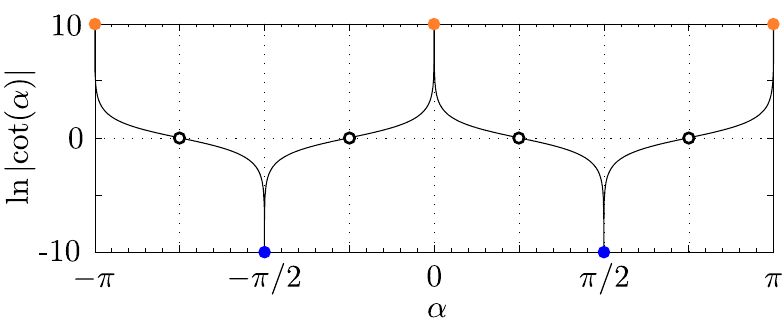}
\caption{\label{locexp}The one-step LE $\lambda^1\equiv\ln|\psi|=\ln|\cot(\alpha)|$ versus the curve slope angle $\alpha$ (see equations (\ref{onestep}) and (\ref{ftleOK})): when the curve is horizontal/vertical ($\alpha=n\pi+0/\tfrac{\pi}{2}$, orange/blue dots) $\lambda^1$ diverges to $+/-\infty$, while when it is orthogonal/parallel to the diagonal ($\alpha=n\tfrac{\pi}{2}+\tfrac{\pi}{4}$, black circles) $\lambda^1$ is zero; this is also the relation between the invariant manifolds slope at fixed points and the full LE (see equation (\ref{stab3})).}
\end{figure}

The last arguments, supported by numerical observation (in figure \ref{diffang1}), a weakly chaotic example), lead to consider a "reduced" expression $\lambda_0^n$ for the FTLE:
\begin{align}
\lambda^n_0\eq\tfrac{1}{n}\sum_{q=0}^{n-1}\ln\left|\psi_q\right|\qd
\label{ftleOK}
\end{align}
which contains only the relevant time-average of the one-step exponent $\lambda^1_q\equiv\ln|\psi_q|$, still preserving the \emph{additive} cocycle property typical of the full FTLE $\chi^n_q$, as in (\ref{ftle}):
\begin{align}
(n+m)\,\lambda^{n+m}_0 \eq n\,\lambda^n_{m}\,+\,m\,\lambda^m_0\qd
\end{align}
The two infinite-time limits $\lambda^\infty_0$ and $\chi^\infty_0$ then coincide and depend only on the ergodic component to which the reference orbit belongs \cite{Pesin}. The main advantage in considering the FTLE expression (\ref{ftleOK}) is its intuitive geometric interpretation: the slopes $\psi$ of the left-invariant manifolds play here the \emph{same} r\^{o}le of the map derivatives in 1D dynamical systems. Thus, the essential difference between 1D and 2D system is the \emph{domain} of the slope: while 1D maps with \emph{bounded derivative} have consequently bounded \emph{slope} angle, this is not true for left-invariant curves, which can bend and turn taking arbitrary angles $\alpha$. Indeed, the local stability represented by the one-step FTLE $\ln|\psi|$ can take values ranging over $\pm\infty$ (see figure \ref{locexp}): this means that points very near to where a left-invariant curve is horizontal/vertical ($\alpha=N\pi+0/\tfrac{\pi}{2}$, orange/blue dots in figure \ref{locexp}) bring respectively very large positive/negative contributions to the FTLE (\ref{ftleOK}), while points near to where a curve is parallel or orthogonal to the diagonal ($\alpha=N\tfrac{\pi}{2}+\tfrac{\pi}{4}$, black circles in figure \ref{locexp}) bring negligible terms. This property is summarized by writing the one-step exponent $\lambda^1\equiv\ln|\psi|$ through the slope angle:
\begin{align}
\lambda^1\equiv\ln|\cot(\alpha)|\eq\ln|\cos(\alpha)|-\ln|\sin(\alpha)|\qc
\label{onestep}
\end{align}
and by studying its zeros and singularities, as depicted in figure \ref{locexp}. In particular, one may wonder how average (\ref{ftleOK}) can converge given the pathologies of function (\ref{onestep}): on the formal side, this is assured by the existence of the Lyapunov exponent for this type of systems (Kingman's theorem \cite{king}); on the geometric side, it is understood by the fact that the singularities of $\ln|\psi|$ along a given curve are always \emph{isolated}, and thus have zero probability also with respect to the measure restricted to such curve.\\
In practice, these considerations induce constraints upon left-invariant curves in the $(x,y)$ representation: the curves cannot be horizontal nor vertical \emph{straight} lines. As a backward check, equation (\ref{stab3}) already brings the very same limitation upon the slopes of the linearised invariant manifolds in the neighbourhood of fixed points.

\subsection{\label{sec:curvat}Scalar curvature evolution}
As much as we addressed the first order properties of left-invariant curves, it is possible to study all the higher orders, obtaining evolutions analogous to (\ref{psievo}); here we consider the second order, to analyse the \emph{curvature} of such curves. To compute it, one has to derive twice with respect to the arc length; by the unit norm of $\dot{\mh{x}}$ and defining the $\frac{\pi}{2}$-rotation $\mh{Y}\defeq\left[\begin{smallmatrix}0&-1\\1&\ \,0\end{smallmatrix}\right]$, such derivative reads:
\begin{align}
\ddot{\mh{x}}\eq\dot{\alpha}\,\mh{Y}\dot{\mh{x}}\qc
	\label{2nd}
\end{align}
implying that $\ddot{\mh{x}}$ has modulus $\dot{\alpha}$ and is perpendicular to $\dot{\mh{x}}$. Inserting (\ref{2nd}) in the definition of curvature $\kappa$ gives:
\begin{align}
\kappa\defeq\frac{\left\|\dot{\mh{x}}\times\ddot{\mh{x}}\right\|}{\|\dot{\mh{x}}\|^3}\ \equiv\ |\dot{\alpha}|\qd
	\label{curv}
\end{align}
To connect curvatures at consequent points of an orbit through an iterative procedure, consider the quantity $\eta$:
\begin{align}
\eta\ \defeq\ \frac{\dot{\psi}}{\dot{y}}\eq-\sigma\,\dot{\alpha}\,(1+\psi)^{\frac{3}{2}}\qsc
\label{eta}
\end{align}
which depends on the arc length derivative of $\psi\equiv\cot(\alpha)$ and thus on $\dot{\alpha}$ itself; the latter is also called the \emph{signed} curvature. Then, by deriving the $\psi$ evolution (\ref{psievo}):
\begin{align}
\dot{\psi}_{n+1}\,\dot{\varphi}_{[n]}\eq f''(x_n)\,\dot{x}_n\ +\ \frac{1}{\psi_n^2}\,\dot{\psi}_n\ \ ,
\label{eta2}
\end{align}
we exploit expression (\ref{phi}) for the expanding factor $\dot{\varphi}$ and then divide both sides of (\ref{eta2}) by $\dot{y}_n\psi_n\equiv\dot{x}_n$, obtaining the desired recursive relation for $\eta$:
\begin{align}
\eta_{n+1}\eq f''(x_n)\ +\ \frac{1}{\psi_n^3}\,\eta_n\qd
\label{eta3}
\end{align}
This map is nonlinear, with a structure similar to the evolution (\ref{psievo}) for $\psi$, as can be seen by re-writing it as:
\begin{align}
\psi_{n+1}\eq f'(x_n)\ -\ \frac{1}{\psi_n^2}\,\psi_n\qd
\label{eta4}
\end{align}
While evolution (\ref{eta4}) is driven by the orbit through the term $f'(x_n)$, (\ref{eta3}) is now driven by  $f''(x_n)$ and by $\psi_n$ itself: thus, the evolutions for $x$, $\psi$ and $\eta$ should be performed in sequence. Such procedure then gives access to the curves slope angle $\alpha\equiv\tn{arccot}(\psi)$ and curvature $\kappa\equiv|\dot{\alpha}|$ by the inverse of relation (\ref{eta}):
\begin{align}
\kappa\eq\frac{|\eta|}{(1+\psi)^{\frac{3}{2}}}\qd
\label{eta5}
\end{align}
Notice that, in order to recover the signed curvature $\dot{\alpha}$, one would need also the logical function $\sigma$, which then requires the evolution of map (\ref{sigevo}); this can be avoided by restricting the analysis to the logarithm of curvatures: this is best suited to probe the \emph{flatness} of curves, due to the very wide range of $\kappa$ values (see Figures \ref{pkt1}, \ref{pkt2}, \ref{TPE}, \ref{TPE2}).

\subsection{\label{splitang}Splitting angles}
As introduced in section \ref{slmap}, reversibility of map (\ref{stan}) by reflection about the diagonal relates stable/unstable curves (see equation (\ref{revcov})). This, in turn, transfers upon slopes $\psi$:
\begin{align}
	\psi^+(x,y)\eq\frac{1}{\psi^-(y,x)}\qc
	\label{sym}
\end{align}
as $\psi^\pm(\mh{x})$ corresponds respectively to the curve $\mathcal{P}^\pm(\mh{x})$. By definition, the splitting angle $\theta=(\alpha^--\alpha^+)$ reads:
\begin{align}
\cot(\theta)\eq\frac{\psi^+\psi^-+1}{\psi^+-\psi^-}\qc
\label{the0}
\end{align}
which, by making use of (\ref{sym}), becomes:
\begin{align}
\cot(\theta(x,y))\eq\frac{\psi^+(x,y)+\psi^+(y,x)}{\psi^+(x,y)\psi^+(y,x)-1}\qd
\label{the}
\end{align}
This shows that $\theta(x,y)$ is symmetric about the diagonal and it can be expressed by the unstable slope $\psi^+$. By coarse graining phase-space into a square grid, it is possible to collect $\psi^+(x,y)$ only, average it in each cell and produce the matrix ${\psi}^+(x_j,y_k)$ whose indices $j$, $k$ are cells positions; the corresponding matrix for $\psi^+(y,x)$ then comes by transposition of the first, allowing to calculate the (grid-averaged) splitting angle by (\ref{the}) without computing the stable slope $\psi^-$.

\subsection{\label{numcon}Algorithm}
We now summarize the algorithm employed in all our numerical analysis; the forward iteration of length $T$:
\begin{align}
\begin{array}{l}
\texttt{for}\quad n=1:T\\
\qquad\qquad x_{n+1}\eq f(x_n)\ -\ x_{n-1}\\
\ \\
\qquad\qquad \psi^+_{n+1}\eq f'(x_n)\ -\ 1/\psi^+_n\\
\ \\
\qquad\qquad \eta^+_{n+1}\eq f''(x_n)\ +\ \eta^+_n/(\psi^+_n)^3\\
\texttt{end}
\end{array}
\end{align}
gives the orbit $x$, the unstable slope $\psi^+$ and the quantity $\eta^+$ to compute the unstable curvature by equation (\ref{eta5}); the backward iteration:
\begin{align}
\begin{array}{l}
\texttt{for}\quad n=T:2\\
\qquad\qquad \psi^-_{n-1}\eq 1/(\,f'(x_{n-1})\ -\ \psi^-_n\,)\\
\ \\
\qquad\qquad \eta^-_{n-1}\eq (\,-f''(x_{n-1})\ +\ \eta^-_n\,)(\psi^+_{n-1})^3\\
\texttt{end}
\end{array}
\end{align}
gives the stable slope $\psi^-$ and the quantity $\eta^-$ for the stable curvature. The last backward iteration essentially allows to check for the symmetry properties between stable and unstable manifolds and to compute the splitting angle $\theta$ as in (\ref{the0}) \emph{point-wise} along an orbit. This is the best way to collect the statistics of splitting angles; as pointed out above, to produce phase-space pictures it is possible to replace expression (\ref{the0}) with (\ref{the}) and avoid the computation of $\psi^-$.

\begin{figure}[t!]
\includegraphics[width=.48\textwidth]{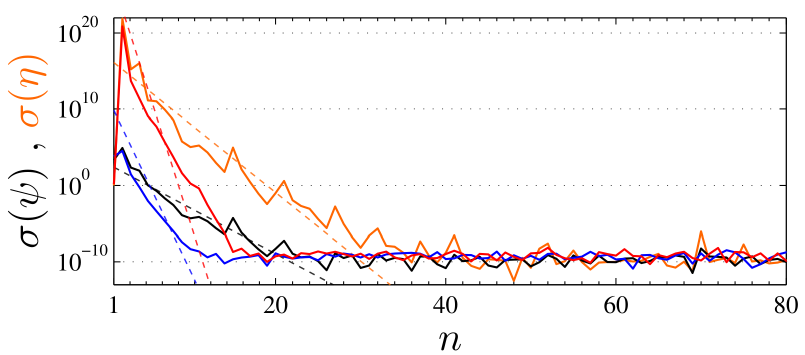}
\caption{\label{psiconv}Exponential decays of the variances of ensembles of $10^7$ random initial conditions $\psi_0$, $\eta_0$ for the evolution (\ref{psievo}) for $\psi$ (in black/blue) and (\ref{eta3}) for $\eta$ (in orange/red); the ensembles run along \emph{fixed} chaotic orbits (initial conditions $(x,y)=(1,2)\times10^{-3}$) for the CT standard map (\ref{CT}) at parameters $K=\pi/2$ (weak chaos, black/orange) and $K=2\pi$ (strong chaos, blue/red). Dashed lines are reference exponentials with twice (in black/blue) and thrice (in orange/red) the FTLE of the reference orbits at finite-time $10^7$; this confirms relations (\ref{psiexp}), (\ref{etaexp}) for the decay of $\delta\psi$ and $\delta\eta$. Decay oscillations are due to the very short finite-times of convergence.}
\end{figure}

\subsubsection{\label{con}Convergence}
\begin{figure}[b!]
\includegraphics[width=.48\textwidth]{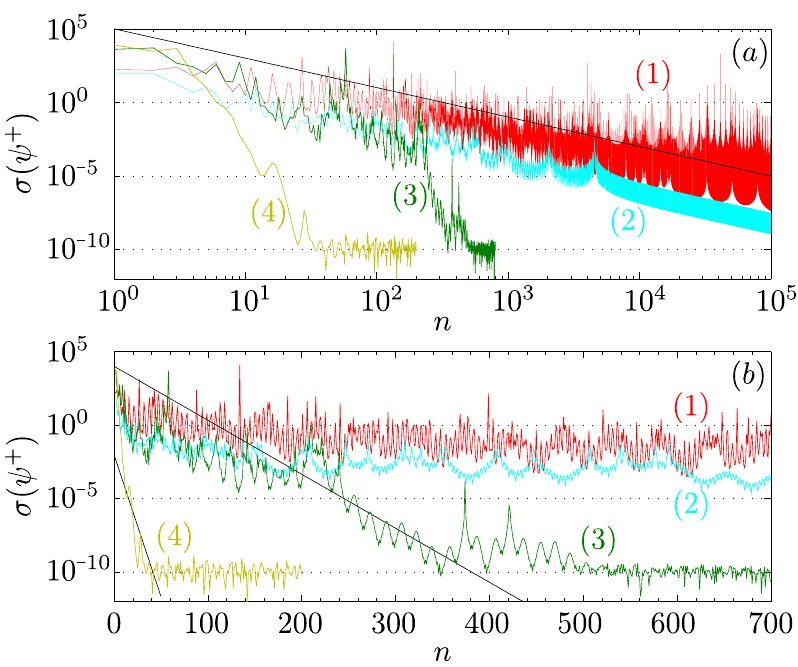}
\caption{\label{comp1}Decays of the variances of ensembles of $10^7$ random initial conditions $\psi_0$ for the $\psi$ evolution (\ref{psievo}) (panel (a): log-log, panel (b): lin-log axis) for four orbits of the CT map (\ref{CT}) at parameter $K=\pi/4$: (1)/(2) open/closed quasi-periodic cycles, (3)/(4) chaotic orbits about period 2 / period 1 unstable fixed points. Each orbit has own decay rate of slope perturbations $\delta\psi$, corresponding to the square of the orbit's perturbations decay: for (1), (2) it is power-law (panel (a), black line is $1/n^2$), while for (3), (4) it is exponential (panel (b), black lines are $\tn{e}^{-2\lambda n}$, with a different exponent $\lambda$ for each chaotic orbit, see equation (\ref{psiexp})).}
\end{figure}

Evolutions (\ref{psievo}) and (\ref{eta3}) are non-autonomous dynamical systems on their own, so it is important to address also their own stability properties, i.e. the sensitivity upon perturbations of \emph{their own} initial conditions $\psi_0$, $\eta_0$. To study both systems at once, we recast them into:
\begin{align}
\left[\begin{matrix}
\psi\\
\eta
\end{matrix}\right]_{n+1}
\,=\,
\left[\begin{matrix}
f'\\
f''
\end{matrix}\right](x_n)
\ +\ 
\left[\begin{matrix}
-\psi^{-2}&0\\
0&\psi^{-3}
\end{matrix}\right]_n\,
\left[\begin{matrix}
\psi\\
\eta
\end{matrix}\right]_n
\label{perew}
\end{align}
This form shows the driving action induced by $f'$, $f''$ and the scaling dependent on $\psi$; by fixing the reference orbit $\{x_n\}$ (and thus the two sequences $\{f'_n\}$, $\{f''_n\}$) the only perturbations in the linearisation of mapping (\ref{perew}) are $\delta\psi$ and $\delta\eta$, which then obey the following evolution:
\begin{align}
\left[\begin{matrix}
\delta\psi\\
\delta\eta
\end{matrix}\right]_{n+1}
\eq
\left[\begin{matrix}-\psi^{-2}&0\\
-3\,\eta\,\psi^{-4}&\psi^{-3}
\end{matrix}\right]_n\,
\left[\begin{matrix}
\delta\psi\\
\delta\eta\end{matrix}\right]_n\qd
\label{staevo}
\end{align}
This is a linear map represented by a lower triangular matrix, which then implies that the linear map from point $0$ to point $n$ (\textit{i.e.} a product of $n$ matrices as (\ref{staevo})) is also lower triangular and, thus, its eigenvalues coincide with its diagonal entries. These are, in modulus, the products of $|\psi_q|^{-2}$ and $|\psi_q|^{-3}$ for $0\leq q<n$ and, by definition (\ref{ftleOK}), they are exponentials of the FTLE $\lambda_0^n$ at time $n$:
\begin{align}
\prod_{q=0}^{n-1}|\psi_q|^\gamma\ \equiv\ e^{\gamma n\lambda_0^n}\qcq\forall\ \gamma\qd
\label{psiprod}
\end{align}
By combining relations (\ref{ktang}) and (\ref{phi}), we notice that the product of subsequent $|\psi|$ along \emph{any type} corresponds to its \emph{expansivity} in tangent space: for chaotic orbits, such product is \emph{exponential} in time (as in (\ref{psiprod})) while for regular orbits it is at most \emph{polynomial}. This implies that, for any chaotic orbit, the stability exponents $\lambda_\psi$ and $\lambda_\eta$ are both {proportional} to the orbit Lyapunov exponent $\lambda$:
\begin{align}
&\lambda_\psi\ \equiv\ \lim_{n\rightarrow\infty}\tfrac{1}{n}\ln\left|\frac{\delta\psi_n}{\delta\psi_0}\right|\ \equiv\ -2\lambda\label{psiexp}\qc\\
&\lambda_\eta\ \equiv\ \lim_{n\rightarrow\infty}\tfrac{1}{n}\ln\left|\frac{\delta\eta_n}{\delta\eta_0}\right|\ \equiv\ -3\lambda\qd\label{etaexp}
\end{align}
This is confirmed in figure \ref{psiconv} for the CT standard map (\ref{CT}), both in the weakly ($K=\pi/2$) and strongly chaotic regime ($K=2\pi$), by the decay of the \emph{variances} of ensembles of initial conditions for both $\psi$ and $\eta$. In the quasi-periodic case, the above exponents are zero but the perturbations $\delta\psi$, $\delta\eta$ also decay respectively as the second and third inverse power of the orbit's perturbations growth. To check this, the same computation of figure \ref{psiconv} is performed in figure \ref{comp1} for $\delta\psi$ in a very weakly chaotic regime ($K=\pi/4$): since the two regular orbits (1) and (2) (respectively open and closed cycles) have linear expansion of perturbations $\sim n$, the corresponding $\psi$ ensembles variance decays as $1/n^2$ (panel (a), log-log). Instead, the variances for the chaotic orbits (3) and (4) (respectively about period 2 and period 1 fixed points) decay exponentially, each with twice the exponent of its own reference orbit (panel (b), lin-log).

\begin{figure*}[t!]
\includegraphics[width=1\textwidth]{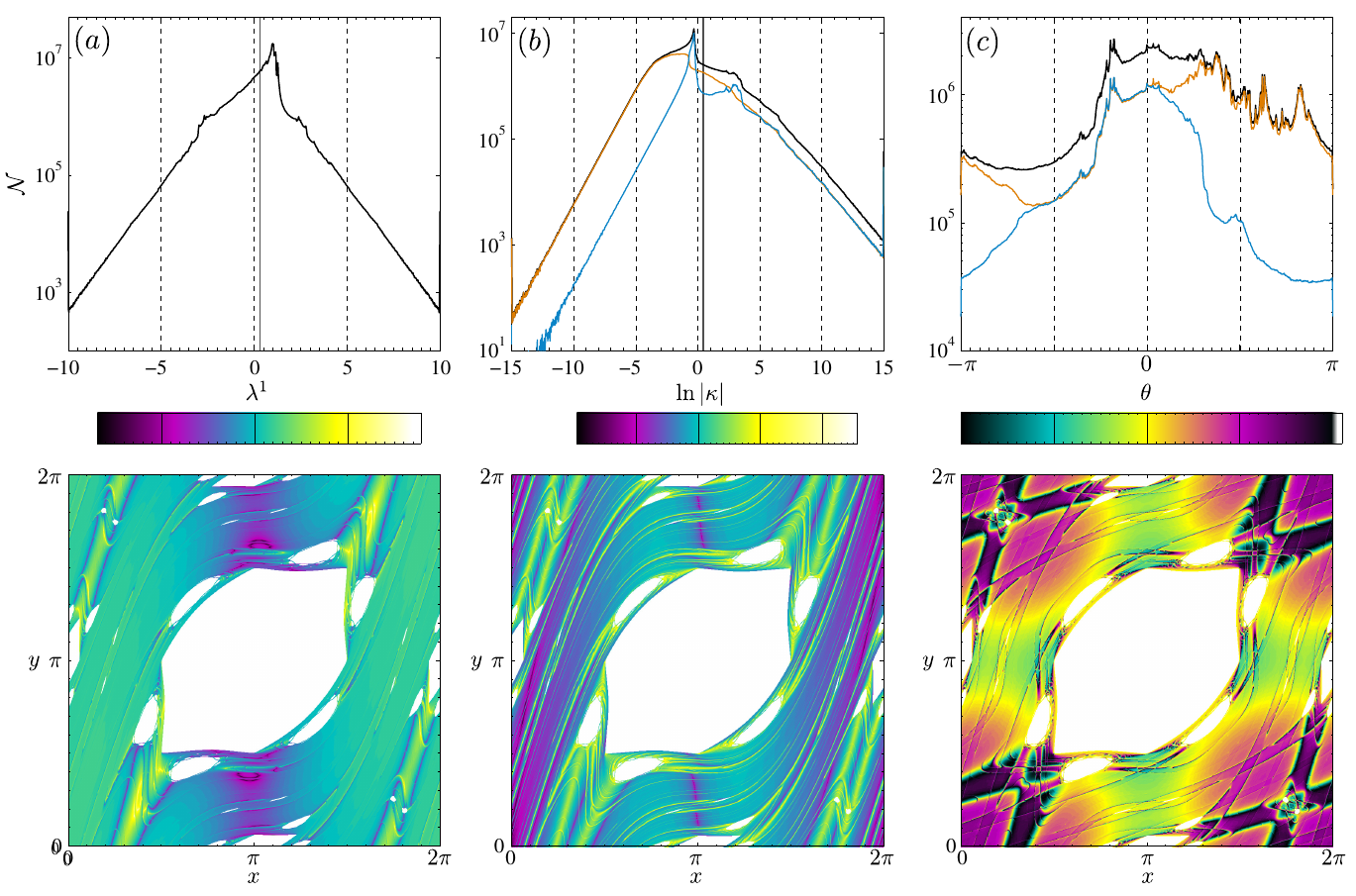}
\caption{\label{pkt1}Each column contains the total (upper panel) and phase-space distributions (lower panel, $10^3\times10^3$ cells) of each function over an orbit of length $10^9$ for the CT map (\ref{CT}) at $K=\pi/2$; color-bars are linked to each total distribution abscissa. Column (a): one-step Lyapunov exponent $\lambda^1=\ln|\psi|=\ln|\cot(\alpha)|$ for the unstable curve; the black vertical line marks the distribution average, which is the orbit's FTLE. Column (b): log-curvature $\ln|\kappa|$ for the unstable curve; the black vertical line marks $\ln|K|$. Column (c): splitting angle $\theta=\alpha^--\alpha^+$. Upper panels (b) and (c): above the total distributions (in black), the conditional distributions for $\ln|\kappa|$ and  $\theta$ as the one-step LE $\lambda^1$ is positive (in orange) and negative (in blue); see also figure \ref{pkt2}.}
\end{figure*}

\begin{figure*}[t!]
\includegraphics[width=1\textwidth]{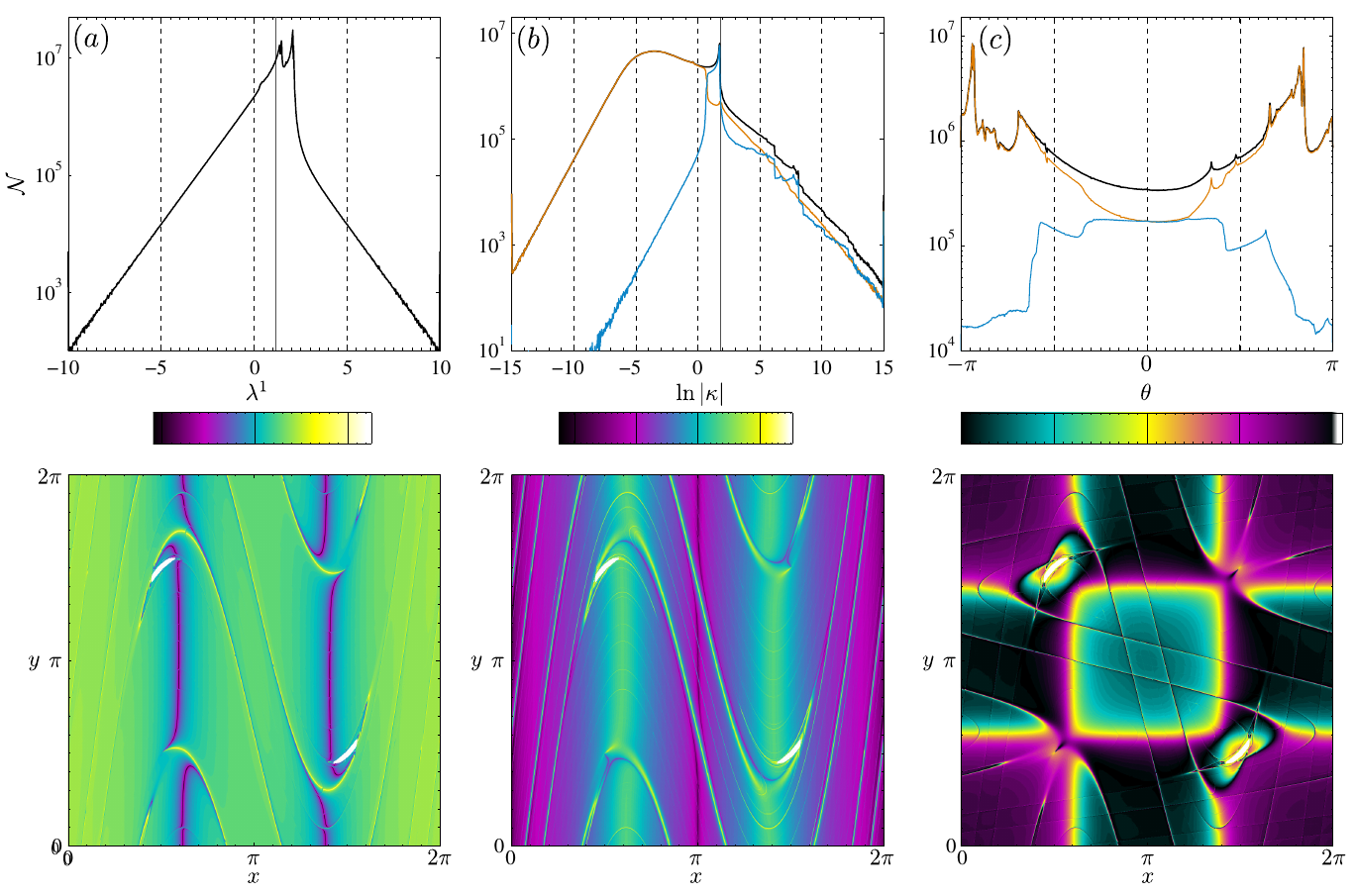}
\caption{\label{pkt2} Same analysis as in figure \ref{pkt1} for the CT map (\ref{CT}) at $K=2\pi$. In the three lower panels, notice that also in this case tiny resonant islands are present (white, unvisited cells), centred around periodic points of period 2 usually called \emph{accelerator modes}. These appear to have no influence on the distributions, since the three observables $\lambda^1$, $\ln|\kappa|$ and $\theta$ are \emph{local} functions. The color-bars for $\lambda^1$ and $\ln|\kappa|$ here are different from figure \ref{pkt1} because of their different probability distribution functions $\mc{N}$.}
\end{figure*}

\begin{figure*}[t!]
\includegraphics[width=.85\textwidth]{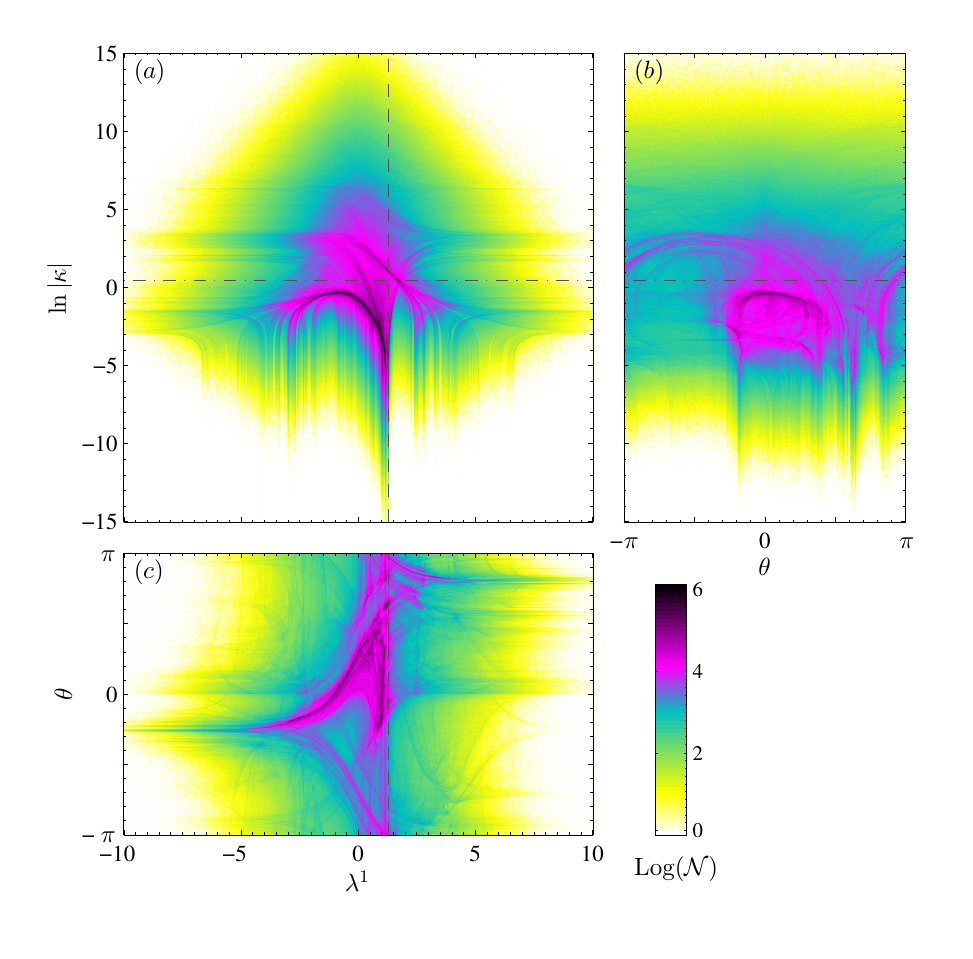}
\caption{\label{TPE}Joint distributions for the pairs $(\lambda^1,\ln|\kappa|)$ (panel (a)), $(\theta,\ln|\kappa|)$ (panel (b)) and $(\lambda^1,\theta)$ (panel (c) computed over the same orbits of length $10^9$ of figure \ref{pkt1} for the CT map (\ref{CT}) at parameter $K=\pi/2$; in color, the $\tn{Log}_{10}$ of the number of events. The lines marks $\lambda^1=\ln|2+K|$ (dashed) and $\ln|\kappa|=\ln|K|$ (dash-dotted), as suggested by figures \ref{YPE}, \ref{YPE2} and expansions (\ref{sols}). }
\end{figure*}

\begin{figure*}[t!]
\includegraphics[width=.85\textwidth]{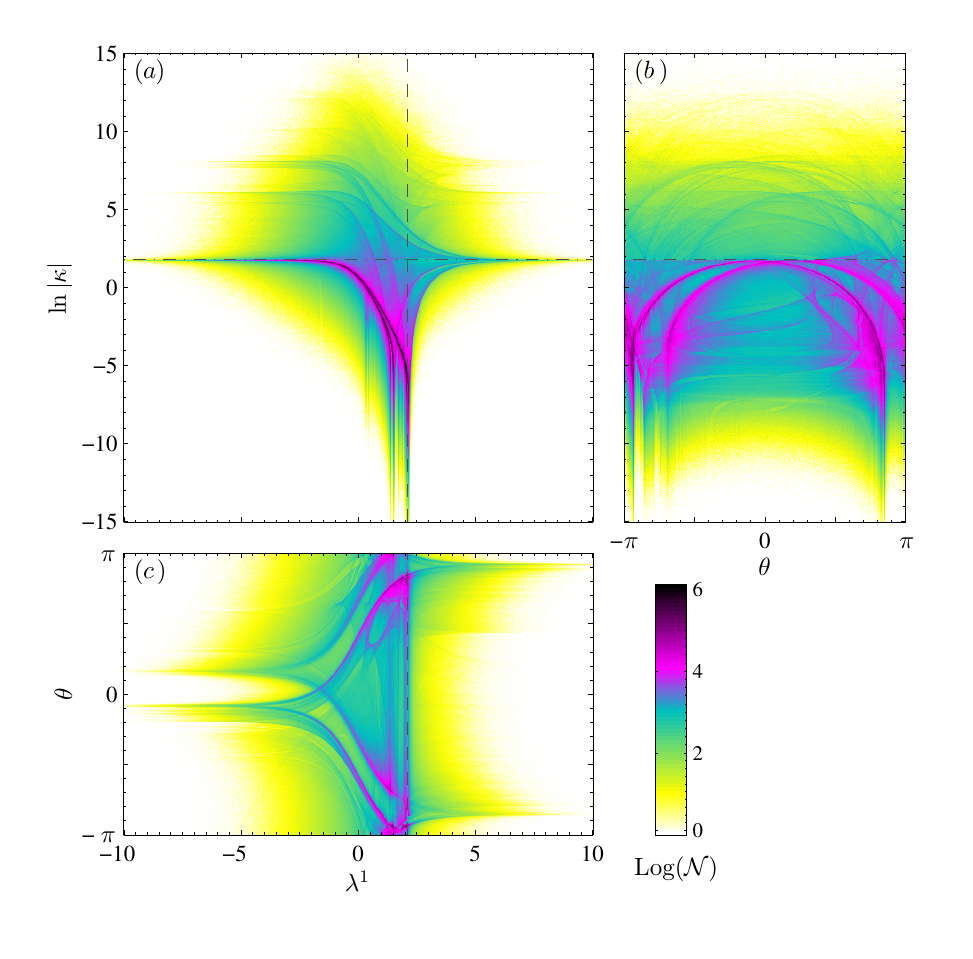}
\caption{\label{TPE2}Same analysis as in figure \ref{TPE} for the same orbits of length $10^9$ of figure \ref{pkt2} for the CT map (\ref{CT}) at parameter $K=2\pi$. Due to higher nonlinearity, the probability is more localized along the lines $\lambda^1=\ln|2+K|$ (dashed), $\ln|\kappa|=\ln|K|$ (dash-dotted), which are the ensemble-averaged leading terms of series (\ref{sols}) (see figures \ref{YPE}, \ref{YPE2}), while the angle $\theta$ is concentrated away from zero.}
\end{figure*}

\begin{figure*}[t!]
\includegraphics[width=.85\textwidth]{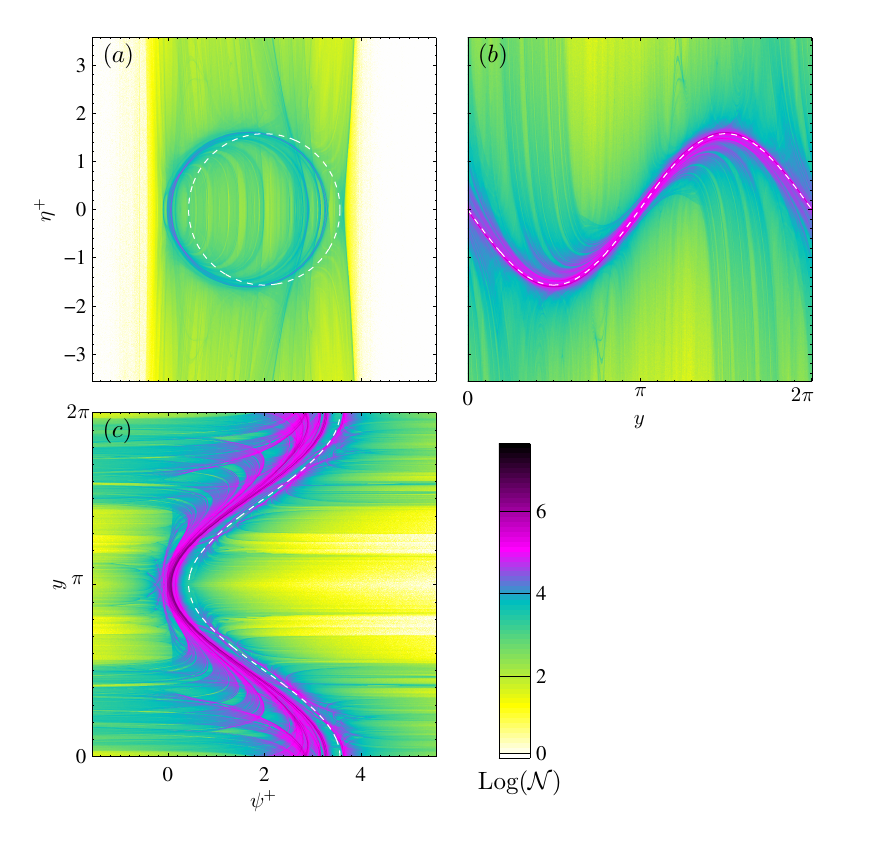}
\caption{\label{YPE}Joint distributions for the pairs $(\psi,\eta)$ (panel (a)), $(y,\eta)$ (panel (b)) and $(\psi,y)$ (panel (c)) calculated over the same data of figure \ref{TPE} for the CT map (\ref{CT}) at parameter $K=\pi/2$. White dashed curves are  $\psi=2+K\cos(y)\equiv f'(y)$ and $\eta=-K\sin(y)\equiv f''(y)$, \textit{i.e.} the first terms of series (\ref{sols}); notice that in panel (a) the $y$ dependence in $\eta$ VS $\psi$ is implicit.}
\end{figure*}

\begin{figure*}[t!]
\includegraphics[width=.85\textwidth]{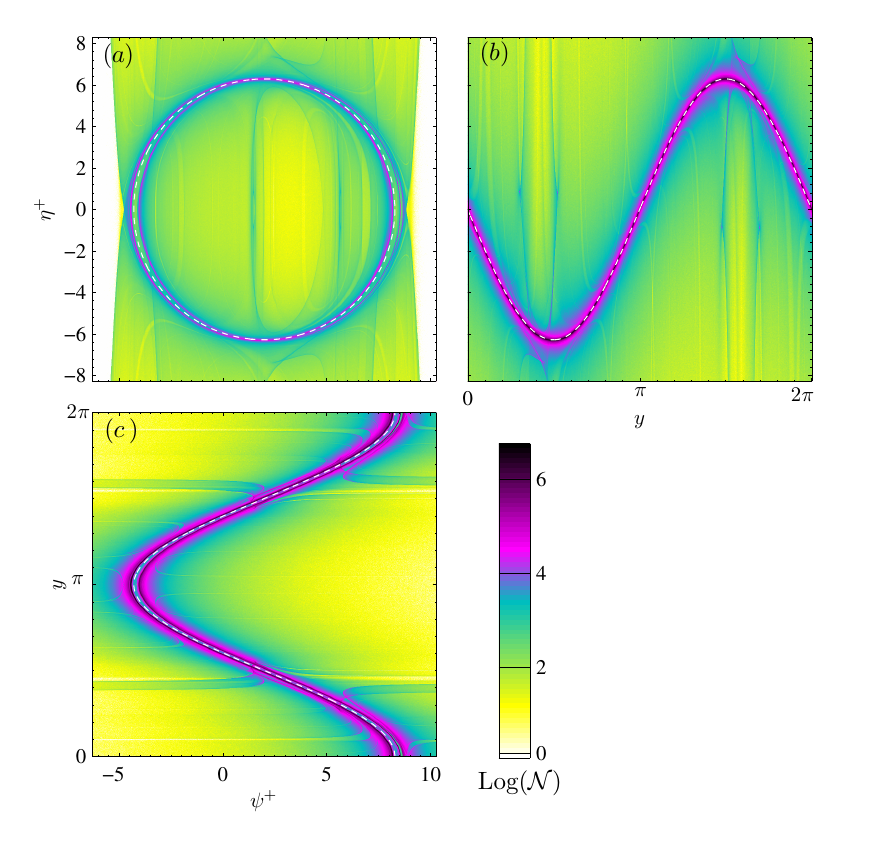}
\caption{\label{YPE2}Same analysis as in figure \ref{YPE} for the same data of figure \ref{TPE2} for the CT map (\ref{CT}) at parameter $K=2\pi$. Due to higher nonlinearity, probability here is localized around the white dashed curves $\psi=2+K\cos(y)\equiv f'(y)$ and $\eta=-K\sin(y)\equiv f''(y)$, \textit{i.e.} the first terms of series (\ref{sols}); this is explained by the faster decay of the series terms, due to the higher value of the FTLEs.}
\end{figure*}

\section{\label{numex}Numerical Experiments}
To exploit the methods illustrated so far, we take the Chirikov-Taylor map (\ref{CT}) at the two values of parameter $K$ already considered in section \ref{numcon}: $K=\pi/2$ (weak chaos)  and $K=2\pi$ (strong chaos), both with chaotic initial conditions $(x,y)=(1,2)\times10^{-3}$, \textit{i.e.} very near the unstable fixed point at the origin. The analysis is applied on three quantities: the one-step Lyapunov exponent $\lambda^1\equiv\ln|\psi^+|$, the log-curvature $\ln|\kappa^+|$ (both for the unstable left-invariant curve) and the splitting angle $\theta\equiv\alpha^--\alpha^+$ between stable/unstable curves. In section \ref{distro}, figures \ref{pkt1} and \ref{pkt2}, we first obtain their probability distributions along with a direct phase-space visualisation: while the former accounts for their global behaviour, the latter allows to picture the point-wise dependence on coordinates. Since, among the three observables, the one-step Lyapunov exponent $\lambda^1$ has the most clear interpretation (see section \ref{geostab}), we calculate also the \emph{conditional} probabilities for both $\ln|\kappa|$ and $\theta$ with respect to both positive and negative values of the one-step LE; in this way we probe which values of curvature and splitting angle are mostly associated to each type of stability. This opens the way to a more detailed investigation through the use of three \emph{joint} distribution functions, one for each pair of observables: $(\lambda^1,\theta)$, $(\lambda^1,\ln|\kappa|)$ and $(\theta,\ln|\kappa|)$ (section \ref{joint}, figure \ref{TPE} and \ref{TPE2}). Although less immediate to interpret, this kind of statistics highlights highly non-trivial dependencies between the three quantities. In doing this, we confirm and extend the results already obtained by \cite{cerb}, \cite{thiff1} in the case of autonomous flows and by \cite{Politi} in the study of the dissipative H\'enon map. In particular, the extremely marked relation between $\lambda^1\equiv\ln|\psi|$ and $\ln|\kappa|$ (already noticed in \cite{thiff3}) hints to search for a more clear view by performing auxiliary joint statistics upon the primal quantities $\psi$, $\eta$ and the second coordinate $y$ (section \ref{funrel}, figures \ref{YPE}, \ref{YPE2}). By simple arguments, we show how the first two quantities can be expanded in function of the third; at once, this explains the relations between local stability $\lambda^1$ and curvature $\kappa$.

\subsection{\label{distro}Distributions \& Visualisations}
The numerical information produced by evolutions (\ref{psievo}) and (\ref{eta3}) becomes a reliable approximation of the true functions $\psi(\mh{x})$, $\eta(\mh{x})$ only after the initial transient of convergence which, therefore, has to be discarded from the statistics of $\lambda^1$ and $\ln|\kappa|$; as shown in figure \ref{psiconv}, the transients depend on the FTLE of each orbit and, for both cases under analysis, they are less than a hundred iterations long. The splitting angle $\theta$ is calculated from (\ref{the0}) by both the stable $\psi^-$ and unstable $\psi^+$ slopes and, since $\psi^-$ is obtained by running evolution (\ref{psievo}) \emph{backward}, both the initial/final (un/stable) transients are dropped. By comparison between figures \ref{pkt1} and \ref{pkt2} (upper panels), we first observe that in both cases the distributions of the one-step exponent and the log-curvature are not bounded: they have exponential tails which are symmetric for $\lambda^1$ and asymmetric for $\ln|\kappa|$; then the two $K$ values can be separated by three main observations:
\begin{itemize}
\item for the one-step exponent, the distribution is far more symmetric about zero for $K=\pi/2$ than for $K=2\pi$, leading to a huge difference between their averages, \textit{i.e.} their FTLEs;  these are respectively $\lambda^T_{[\pi/2]}=0.298$ and $\lambda^T_{[2\pi]}=1.172$ at time $T=10^9$ (figures \ref{pkt1}, \ref{pkt2}, upper panels (a), black vertical line).
\item for the log-curvature, at $K=\pi/2$ the positive tail is longer than the negative one, while at $K=2\pi$ it is the opposite; the conditional distributions (upper panels (b), orange/blue graphs) show that the majority of points having large \emph{negative} $\ln|\kappa|$ also have \emph{positive} one-step exponents, suggesting that local instability is concentrated in regions where the curves are almost \emph{flat}.
\item for the splitting angle distributions, the differences between the two cases are stronger since $\theta=0$ is a relative maximum (inside a wide plateau) for $K=\pi/2$ and an absolute minimum for $K=2\pi$; interestingly, in both cases the probability of null splitting angle is equally divided between points with positive/negative one-step exponents (upper panels (c), orange/blue graphs), while most of the points with positive $\lambda^1$ also exhibit $|\theta|>\pi/4$.
\end{itemize}
Notice that the illustrated features have minor details that vary between the two dynamical cases: the tails of negative log-curvature for the orange ($\lambda^1>0$) and blue ($\lambda^1<0$) graphs differ by almost two orders of magnitudes at $K=\pi/2$ and by more than four orders at $K=2\pi$, while such difference is almost zero for both positive tails. On the same line, the range of $\theta$ values that have equal conditional probability (both around $\theta=0$) is much wider for $K=\pi/2$ than that for $K=2\pi$. By consequence, all the above considerations can lead to solid conclusions only about the behaviour of $\lambda^1$: the regions where the one-step exponent is positive have much higher probability to contain curves that are simultaneously flat \emph{and} markedly transversal ($|\theta|>\pi/4$), \textit{i.e.}  that \emph{locally} resemble a \emph{uniformly} hyperbolic system. 
\subsubsection{\label{tang}Tangencies \& Partitions}
Finally, it should be noted that the phase-space structures in figures \ref{pkt1} and \ref{pkt2} (lower-panel (c)), highlighted by the splitting angle values, are directly related to the shape of the \emph{generating partition} of the symbolic dynamics associated to the CT map (\ref{CT}); in particular, in figure \ref{pkt2}, the iso-curves at null splitting angle, which identify the family of homoclinic tangencies, correspond exactly to the generating partition curves depicted in figure 3 in \cite{chris1} and figure 7 in \cite{chris2}, where they were found by a different method. A non-trivial feature in identifying the tangencies by the present phase-space plots comes from the extreme compression of some of the splitting angle iso-curve: in the upper-left region of the same panel as above, one can clearly see an almost vertical curve of tangencies that bends toward right and continue into a diagonal, very narrow scar; the latter thus also seems to be part of the partition border. This is confirmed by the method employed in \cite{chris1} (figure 3) but, because of the averaging procedure in each cell, it is not at all clear from the present phase-space picture, suggesting that the analysis is not optimal for finding precise partitions. Interestingly, on the other hand, here the very narrow scars appear to be related to the location of \emph{avoided crossings}\cite{chris2} in the partition boundary. We do not address such problem here but remark that, in general, the phase-space visualization of the splitting angle already gives a global picture of how the associated generating partition looks like.

\subsection{\label{joint}Joint Distributions}
In spite of the complex structure of mixed phase-spaces, the results above suggest to search for specific relations between the three observables under study; to this end, we consider the joint distributions of events for each pair $(\lambda^1,\theta)$, $(\lambda^1,\ln|\kappa|)$ and $(\theta,\ln|\kappa|)$ over a regular grid of $10^3\times 10^3$ bins (figures \ref{TPE}, \ref{TPE2}). In this respect, the conditional distributions for $\ln|\kappa|$ and $\theta$ correspond to the integration of the joint distributions for $(\lambda^1,\theta)$ and $(\lambda^1,\ln|\kappa|)$ over positive/negative values of $\lambda^1$  (respectively orange/blue graphs in figures \ref{pkt1}, \ref{pkt2}, upper panels (b), (c)); therefore, the 2D histograms information of figures \ref{TPE}, \ref{TPE2} extends what has been already discussed by conditional distributions: the color represents the logarithm (in base $10$) of the number of collected events for each pair of observables. For both values of $K$, this reveals the coexistence of very sharp trends along with large deviations, explaining why the conditional probabilities for $\ln|\kappa|$ and $\theta$ exhibit both separation and equipartition; once the 2D histograms are integrated over $\lambda^1\gtrless0$, the accumulations of events (black/purple bins in figures \ref{TPE}, \ref{TPE2}, panels (a), (c)) induce separation or equipartition depending on their main position with respect to $\lambda^1=0$. Indeed, at $K=\pi/2$ the accumulations largely cross such value exactly in the ranges $\sim\theta\in[-\pi/4,0]$ and $\sim\ln|\kappa|\in[-1,\infty]$, where the conditional distributions are \emph{equal} (compare to figure \ref{pkt1}, upper panels (b), (c)). Instead, at $K=2\pi$ the accumulations stay in the positive $\lambda^1$ semi-plane for $\sim|\theta|\in[\pi/4,\pi/2] $ and $\sim\ln|\kappa|\in[-\infty,1]$, leading to the high \emph{separation} of conditional distributions (compare to figure \ref{pkt2}, upper panels (b), (c)). The differences between the two cases can be ascribed to the more regular structure of curves at $K=2\pi$ and to the high number of resonant islands at $K=\pi/2$ which brings rare events with high curvatures and small splitting angles; again, the 2D distributions confirm the connection between higher local instability and the locally hyperbolic character of left-invariant curves. What remains to be explained is the extremely sharp relation between $\lambda^1$ and $\ln|\kappa|$; this is addressed in the next section, showing how such feature essentially depends on the nonlinearity strength $K$.

\subsection{\label{funrel}Functional Relations}
The relation between one-step exponent $\lambda^1\equiv\ln|\psi|$ and log-curvature $\ln|\kappa|$ (figures \ref{TPE}, \ref{TPE2}, panels (a)) exhibits a coincidence between the accumulation of events (black/purple bins) and the values $\lambda^1=\ln|K+2|$ (dashed vertical line) and $\ln|\kappa|=\ln|K|$ (dash-dotted horizontal line), especially for $K=2\pi$. In trying to understand such behaviour, two considerations come in help:
\begin{itemize}
\item from equation (\ref{eta5}), curvature $\kappa$ and slope $\psi$ are related through the auxiliary quantity $\eta$ ;
\item from definition (\ref{psi}) for $\psi$ and (\ref{eta}) for $\eta$, these can be interpreted as first and second derivatives of a \emph{local representation} $x(y)$ for the left-invariant curve:
\begin{align}
\psi=\frac{\dot{x}}{\dot{y}}\sim\left.\frac{\tn{d}x}{\tn{d}y}\right|_{y}\qcq\eta=\frac{\dot{\psi}}{\dot{y}}\sim\frac{\tn{d}\psi}{\tn{d}y}\sim\left.\frac{\tn{d}^2x}{\tn{d}y^2}\right|_{y}\ 
\label{interp}
\end{align}
\end{itemize}
Following such hints, along with the low magnitude of $K$ with respect to the range of $\psi$ and $\kappa$, we study the pair $(\psi,\eta)$ avoiding the logarithms and ignoring their large deviations; in figures \ref{YPE}, \ref{YPE2}, panels (a) and (d), this reveals an accumulation of events around a precise relation between $\psi$ and $\eta$: the majority of points falls upon a circle of radius $K$ centred in $(\psi,\eta)=(2,0)$ (white dashed curve). Moreover, in the same figure, panels (b), (c) and (e), (f), the same analysis for $(\psi,y)$ and $(y,\eta)$ confirms the interpretation pictured in (\ref{interp}) exactly: both $\psi$ and $\eta$ exhibit a marked statistical dependence upon $y$.

\subsubsection{Implicit Solutions}
To understand pictures \ref{TPE}-\ref{YPE2}, we need to identify relations between $\psi$, $\eta$ and $y$.
Evolution form (\ref{perew}) allows to write its \emph{formal} solutions by the definition of matrices:
\begin{align}
\mh{\Gamma}_j^n\defeq\prod_{q=j}^{j+n-1}
\left[\begin{matrix}
\tn{-}\psi^{{-}2}&0\\
0&\psi^{{-}3}
\end{matrix}\right]_{q}
\equiv
\left[\begin{matrix}
(\tn{-}1)^n\tn{e}^{\tn{-}2n\lambda_j^n}&0\\
0&\tn{e}^{\tn{-}3n\lambda_j^n}
\end{matrix}\right]
\label{peaux}
\end{align}
whose properties derive from the FTLE in formula (\ref{ftleOK}): if the exponent $\lambda_j^n$ is positive, matrix $\mh{\Gamma}_j^n$ decays to zero for $n\rightarrow\infty$. Indeed, by inserting (\ref{peaux}) into formal summations of evolution (\ref{perew}) and using the fact that $y_q=x_{q-1}$, we can obtain its finite-time solutions:
\begin{align}
\left[\begin{matrix}
\psi\\
\eta
\end{matrix}\right]_{n}
\eq
\mh{\Gamma}_{0}^{n}
\left[\begin{matrix}
\psi\\
\eta
\end{matrix}\right]_0
\ +\ 
\sum_{q=1}^{n}\mh{\Gamma}_{q}^{n-q}
\left[\begin{matrix}
f'\\
f''
\end{matrix}\right](y_{q})\qd
\label{pesol1}
\end{align}
As said, $\mh{\Gamma}^n_0$ decays to zero for chaotic orbits, allowing to drop the term ${[\psi,\eta]_0}$; by \emph{shifting} equation (\ref{pesol1}) from $\mh{x}_0\equiv\mh{x}$ to $\mh{x}_{-n}$ and from $q$ to $q-n$ leaves ${[\psi,\eta]_0}\equiv[\psi,\eta](\mh{x})$ on the left hand side for \emph{any} choice of $n$, yielding:
\begin{align}
\left[\begin{matrix}
\psi\\
\eta
\end{matrix}\right](\mh{x})
\eq
\sum_{q=0}^{\infty}\mh{\Gamma}^q_{-q}
\left[\begin{matrix}
f'\\
f''
\end{matrix}\right](y_{-q})\qd
\label{pesol3}
\end{align}
Since the $\mh{\Gamma}$ are diagonal, this can be written separately for $\psi$ and $\eta$ and, by expression (\ref{peaux}), through the FTLE:
\begin{align}
&\psi(\mh{x})=\sum_{q=0}^{\infty}(-1)^q\,e^{-2q\lambda^q_{-q}}\,f'(y_{-q})\qc\nonumber\\
&\eta(\mh{x})=\sum_{q=0}^{\infty}\sigma_q\,e^{-3q\lambda^q_{-q}}\,f''(y_{-q})\qd
\label{sols}
\end{align}
These solutions are series running over the negative semi-orbit of $\mh{x}$ whose terms $f'(y_{-q})$, $f''(y_{-q})$ are weighted by factors that decay \emph{exponentially} in $q$; respectively for $\psi$/$\eta$, the decay rate is twice/trice the orbit's FTLE, exactly as for their perturbations (again, see section \ref{numcon}). Thus, solutions (\ref{sols}) explain at once the numerical results of figures \ref{YPE}, \ref{YPE2}: for the CT map (\ref{CT}), the leading terms of the two series are the functions $f'(y)=2+K\cos(y)$ for $\psi$ and $f''(y)=-K\sin(y)$ for $\eta$, represented in panels (b),(e) and (c),(f) respectively (white dashed curves). The two values of the FTLE, $\lambda^T_{[\pi/2]}=0.298$ and $\lambda^T_{[2\pi]}=1.172$, then complete the description:
\begin{itemize}
\item since at $K=\pi/2$ the exponent is very small, the distribution in the $(\psi,y)$ plane (panel (c)) deviates from $f'(y)$ much more than at $K=2\pi$ (panel (f));
\item at both values of parameter $K$, the leading term is more important for $\eta$ (panels (b),(e)) than for $\psi$ (panels (c),(f)) because in the former the exponential weights decay faster than in the latter (respectively thrice and twice the orbit's FTLE).
\end{itemize}
It should be remarked that the series in (\ref{sols}) have purely formal meaning since, by the presence of the FTLE's $\lambda_{-q}^q$, they involve the products of function $\psi$ itself evaluated over the negative semi-orbit of point $\mh{x}$. Therefore, the series for $\psi(\mh{x})$ is an implicit expression but, nevertheless, it can still be exploited to construct approximations.
\begin{figure}[t!]
\includegraphics[width=\columnwidth]{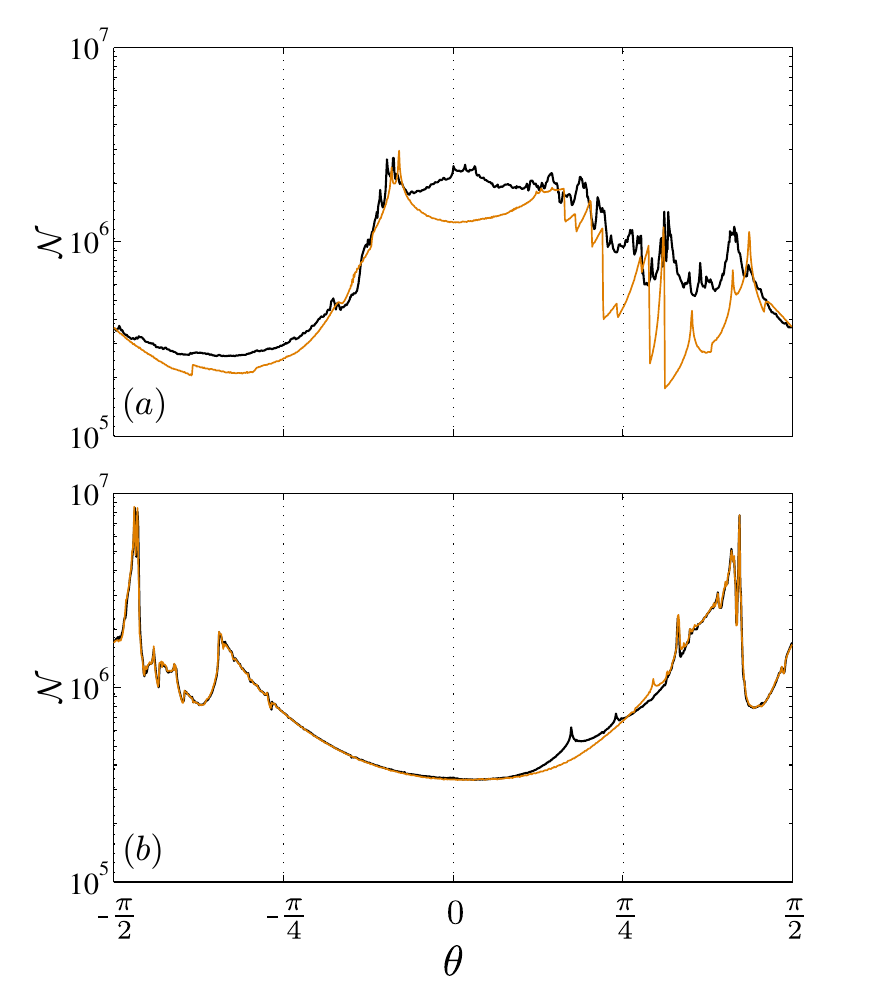}
\caption{\label{app}Probability distribution for the splitting angle $\theta$ calculated by approximation (\ref{appr}) (orange graphs) and by the original data (black graphs) for the CT map (\ref{CT}) at parameter $K=\pi/2$ (panel (a)) and $K=2\pi$ (panel (b)); the black graphs are the same as in the columns (c) of figures \ref{pkt1} and \ref{pkt2}.}
\end{figure}
\subsubsection{\label{appro}Explicit Approximations}
Depending on the FTLE magnitude of the orbit under analysis, it is possible to consider finite \emph{truncations} of the series (\ref{sols}); for function $\psi(\mh{x})$, this produces approximated solutions with errors proportional to the maximum value of $f'$ times the exponential factor of the first discarded term. The fact that series (\ref{sols}) truncated at the $N-th$ term still requires to evaluate $\psi$ itself at the $N$ pre-images of the target point $\mh{x}$ can be overtaken by re-inserting recursively the $N-1$ lower approximations. It is then possible to show that such procedure coincides exactly with the $N$-th truncation of the formal solution:
\begin{align}
&\psi(\mh{x})=f'(y)-\frac{1}{f'(y_{-1})-\frac{1}{f'(y_{-2})-\frac{1}{f'(y_{-3})-\frac{1}{\dots}}}}
\label{cont}
\end{align}
which can be obtained by reinserting the original evolution (\ref{psievo}) into itself; this has the structure of a continued fraction, and by its truncations or, equivalently, of series (\ref{sols}), the lowest order approximations are the same:
\begin{align}
\psi^O(x,y)&\eq f'(y)\ \ ,\nonumber\\
\psi^I(x,y)&\eq f'(y)\ -\ \frac{1}{f'(f(y)\ -\ x)}\ \ ,\label{appr}\\\nonumber
\psi^{II}(x,y)&\eq f'(y)\ -\ \frac{1}{f'(y_{-1})\ -\ \frac{1}{f'(y_{-2})}}\ \ ,
\end{align}
with $y_{-1}=f(y)-x$ and $y_{-2}=f(y_{-1})-y$, as induced by (\ref{inv}) (the ${\tn{mod}\,2\pi}$  for the CT map (\ref{CT}) can be dropped since $f'$ is already periodic). Notice that it is \emph{not} possible to obtain the next higher order approximation from the previous ones, because the additional terms have to be added at lower and lower denominators. In figure \ref{app} we compare the statistics of the splitting angle $\theta$ obtained by second-order approximation $\psi^{II}$ versus the exact distributions (already shown in figures \ref{pkt1} and \ref{pkt2}): while the result for $K=2\pi$ (panel (b)) shows an excellent agreement between the approximated (in orange) and the true data (in black), for $K=\pi/2$ (panel (a)) the two graphs coincide only qualitatively and, in particular, the correct shape of the distribution around $\theta=0$ is missing. This is direct consequence of the much lower values of the FTLEs for $K=\pi/2$, which induce a much slower decay of the weights in series (\ref{sols}); from the FTLE value, one could infer a priori the need for higher order approximations.

\section{Conclusions}
By exploiting the connection between the left-invariant curves and the covariant Lyapunov vectors forming the Oseledets' splitting, we first highlight a direct relation between the one-step Lyapunov exponent and the slope of the curves, which gives a nice interpretation of stability through geometry, paralleling the FTLE calculation in 1D systems. Interestingly, the relation can be extended to any invertible map of the plane. Through the analysis of joint statistics for the one-step exponent, the curvature and the splitting angle between stable/unstable curves, we find definite relations connecting the three quantities which suggest a precise picture: in the regions of phase-space where the one-step exponent is larger and positive (unstable), the curves are flat and the splitting angles are bounded away from zero, that is, the system has a \emph{locally} hyperbolic structure. In particular, this leads to detect very sharp relations between slopes and curvatures which can be understood by expanding them as weighted averages of the first and second derivatives of the map: the leading terms of such series both depend on one of the phase-space coordinates, completely explaining the presence of sharp relations; on the other hand, the large deviations from such average trends can be understood by the rate of decay of the higher-order terms of those same series, which, in turn, explicitly depends on the magnitudes of the finite-time Lyapunov exponents. 

\begin{acknowledgments}
The authors are grateful to the reviewers for pointing out the interesting analogy with billiards and wave-fronts curvature. This work has been partially supported by the MIUR$-$PRIN project `Nonlinearity and disorder in classical and quantum transport' and by the MIUR$-$FIRB project `Futuro in ricerca', number RBFR08UH60.
\end{acknowledgments}

\appendix*
\subsection{\label{appx:a1}M\"obius evolution}
To extend the evolutions (\ref{psievo}) and (\ref{eta3}) to any $\mathcal{C}^2$ map $\mh{\Phi}$ of the plane we make use of the homomorphism between the $SL(2,\mathbb{R})$ group (in which any invertible Jacobian matrix is included by normalisation) and the M\"obius group of linear-fractional transformations. By first rewriting evolution (\ref{curvetan}) for unit tangent vectors:
\begin{align}
\dot{\mh{x}}_{n+1}\,\dot{\varphi}_n\eq\mh{J}_n\,\dot{\mh{x}}_n\qc
\end{align}
we assume $\mh{J}_n$ invertible and $\dot{\mh{x}}$ as in definition (\ref{psi}):
\begin{align}
\mh{J}_n=\begin{bmatrix}A&B\\C&D\end{bmatrix}_n\equiv\left.\frac{\partial\bos{\Phi}}{\partial\mh{x}}\right|_{\mh{x}_n}\ ,\ \dot{\mh{x}}_n=\frac{\sigma_n}{\sqrt{1+\psi^2_n}}\begin{bmatrix}\psi_n\\1\end{bmatrix}\nonumber\qd
\end{align}
Then we get the corresponding evolutions for $\psi$ and $\sigma$:
\begin{align}
&\psi_{n+1}\eq\frac{A_n\psi_n+B_n}{C_n\psi_n+D_n}\qc\label{gpevo}\\
&\sigma_{n+1}\eq\tn{sign}\left(C_n\psi_n+D_n\right)\sigma_n\qd
\label{gevo}
\end{align}
In this setting, the forward/backward FTLEs $(\lambda_0^n)^\pm$ are:
\begin{align}
(\lambda^n_0)^\pm=\tfrac{1}{n}\sum_{q=0}^{n-1}\ln\left|C_q\psi_q^\pm+D_q\right|\quad\to\quad\lambda^\pm\qc
\label{ftle3}
\end{align}
where the same arguments of section \ref{geostab0} are applied to consider the FTLE in a reduced form. Notice that now the slope values for which the one-step exponent $\lambda^1_n\equiv\ln\left|C_n\psi_n+D_n\right|$ is singular are again the horizontal ones ($\psi\equiv\cot(\alpha)\to\pm\infty$) but no more the vertical ones; instead, there are other special slopes:
\begin{align}
\psi_n\eq-\frac{D_n}{C_n}
\end{align}
which induce \emph{local} singularities of the one-step exponents. Again, for regular maps, these are isolated along the curves, thus harmless to the convergence of series (\ref{ftle3}). By a procedure equivalent to \ref{sec:curvat}, also evolution (\ref{eta3}) for $\eta=\dot{\psi}/\dot{y}$ is extended to the general case:
\begin{align}
\eta_{n+1}=\ \frac{1}{\gamma_n^3}\left(\det(\mh{J}_n)\eta_n+\left(a_n+b_n\psi_n+c_n\psi^2_n\right)\right)
\label{geevo}
\end{align}
with $\gamma_n\defeq \left(C_n\psi_n+D_n\right)$ and the coefficients given by:
\begin{align}
&a_n\eq\mh{v}_n\cdot(D\nabla B-B\nabla D)_n\qc\nonumber\\
&b_n\eq\mh{v}_n\cdot(D\nabla A-A\nabla D+C\nabla B-B\nabla C)_n\qc\nonumber\\
&c_n\eq\mh{v}_n\cdot(C\nabla A-A\nabla C)_n\qc\nonumber
\end{align}
and $\mh{v}_n\equiv[\psi_n,1]^T$ , $\nabla$ the gradient operator and $"\cdot"$ the scalar product. The Lyapunov exponents for $\psi$ and $\eta$ are then found by perturbing evolutions (\ref{gpevo}), (\ref{geevo}) keeping fixed all the orbit-dependent coefficients; this leads to express $\lambda_\psi$ and $\lambda_\eta$ by the two orbit's exponents $\lambda^+\geq\lambda^-$:
\begin{align}
&\lambda^\pm_{\psi}\ \equiv\ \lim_{n\rightarrow\pm\infty}\tfrac{1}{n}\ln\left|\delta\psi_{n}\right|\eq-(\lambda^+-\lambda^-)\nonumber\\
&\lambda^\pm_{\eta}\ \equiv\ \lim_{n\rightarrow\pm\infty}\tfrac{1}{n}\ln\left|\delta\eta_{n}\right|\eq-(\lambda^+-\lambda^-)\mp\lambda^\pm
\label{gstab1}
\end{align}
The standard-like case corresponds to have $\gamma_n=\psi_n$ and $\lambda_{\psi}=-2\lambda^+$ , $\lambda_{\eta}=-3\lambda^+$, while $a_n=b_n=0$, $c_n=\psi_nf''_n$. Conditions (\ref{gstab1}) imply that the convergence of $\psi$ only requires the that orbit exponents are non-degenerate, $\lambda^-\neq\lambda^+$ for \emph{any} sign of the exponents, while for $\eta$ three cases appears:
\begin{itemize}
\item $\lambda^-<0<\lambda^+$ (hyperbolic): the same non-degeneracy condition $\lambda^-\neq\lambda^+$ is enough for $\eta^\pm$;
\item $0<\lambda^-<\lambda^+$ (purely expansive): this case requires that $2\lambda^-<\lambda^+$, otherwise only $\eta^+$ converges ;
\item $\lambda^-<\lambda^+<0$ (purely contractive): this case requires that $\lambda^-<2\lambda^+$, otherwise only $\eta^-$ converges;
\end{itemize}
the last two cases mean that, if both the tangent subspaces are expanded/contracted, the deformations should be ``different enough'' to have the convergence of both $\eta^\pm$. 

\subsection{\label{appx:a2}Billiards Analogy}
Here we point out a suggestive analogy between the evolution (\ref{psievo}) for the manifolds' slopes and the evolution of the curvature of a \emph{wave-front} in a bidimensional billiard; consider first the canonical set of coordinates $(x,p)\equiv (x,x-y)$ introduced in (\ref{sta}) and the corresponding manifold' slope $\hat{\psi}$ transformed according to:
\begin{align}
\hat{\psi}\ \defeq\ \frac{\dot{p}}{\dot{x}}\eq1\ -\ \frac{1}{\psi}\qd
\label{trapsi}
\end{align}
Notice that, if the time-period of the delta-kicks is set to $\tau$ instead of unity, map (\ref{sta}) can be re-written in the form:
\begin{align}
\left(\begin{array}{c}
p'\\
x'
\end{array}\right)
\eq
\left(\begin{array}{c}
p\ +\ \tau F(x)\\
x\ +\ \tau p\ +\ \tau^2 F(x)
\end{array}\right)
\label{stanew}
\end{align}
In the framework of canonical transformations, one may consider a generating function $S(x)$ and write down the momentum $p$ as its derivative with respect to $x$, so that:
\begin{align}
p\eq\frac{\partial S}{\partial x}(x)\quad\Rightarrow\quad\hat{\psi}\eq\frac{\dot{p}}{\dot{x}}\equiv\frac{\partial^2 S}{\partial x^2}(x)\qc
\label{trapsi2}
\end{align}
and, interestingly, the manifold slope $\hat{\psi}$ corresponds to the second derivative of the generating function $S$, so is proportional to its \emph{curvature}. The last step is to derive the evolution for the slope $\hat{\psi}$ by \textit{e.g.} equation (\ref{gpevo}), by first writing the Jacobian matrix of map (\ref{stanew}):
\begin{align}
&\mh{J}_n\eq\left(\begin{array}{cc}
1 & \tau F'(x_n)\\
\tau & 1+\tau^2 F'(x_n)
\end{array}\right)\qc
\label{trapjac}
\end{align}
obtaining, after some simple re-adjustment:
\begin{align}
\hat{\psi}_{n+1}\eq\frac{1}{\tau\ +\ \cfrac{1}{\hat{\psi}_{n} + \tau F'(x_n)}}\qd
\label{trapevo}
\end{align}
Quite remarkably, evolution (\ref{trapevo}) exactly coincides with the evolution \cite{garr,cher} of the curvature of a wave-front at a point which freely propagates for a time $\tau$ after reflecting perpendicularly on a boundary with curvature $K=\frac{\tau}{2}F'(x)$; by setting $F(x)\equiv-V'(x)$ for some choice of kicked potential $V(x)$, this becomes $K=-\frac{\tau}{2}V''(x)$, i.e. the boundary curvature is proportional to the curvature of the potential function. We are not aware of any interpretation of map (\ref{stanew}) as billiard dynamics, although the latter has an Hamiltonian description. It should be noted that, in a billiard, $\tau$ is non-constant and depends on the trajectory; instead, in the delta-kicked model, $\tau$ is fixed, even if not chosen constant. A possible interpretation of such analogy may go like this: once the potential $V(x)$ is switched on, the momentum $p$ is istantaneously changed to $p'=p+\tau F(x)$; after that, the point $x$ propagates freely for a time $\tau$ at constant speed $p'$. Parallel to this, one can imagine an associated curve $S(x)$ (the wave-front in the billiard or the generating function in the canonical picture) whose \emph{radius of curvature} is first scattered (by \emph{reflection} or \emph{refraction}) through the potential curvature and then grows linearly in time for a period $\tau$. After a single cycle, such compound evolution would correspond to (\ref{trapevo}). The connection between manifolds' slopes and the Hessian of the generating function will be addressed in a future work.


\begin{thebibliography}{25}
\providecommand{\natexlab}[1]{#1}
\providecommand{\url}[1]{\texttt{#1}}
\expandafter\ifx\csname urlstyle\endcsname\relax
  \providecommand{\doi}[1]{doi: #1}\else
  \providecommand{\doi}{doi: \begingroup \urlstyle{rm}\Url}\fi

\bibitem[C.L.~Wolfe(2007)]{Wolfe}
R.M.~Samelson and C.L.~Wolfe.
\newblock An efficient method for recovering Lyapunov vectors from singular
  vectors.
\newblock \emph{Tellus A}, 59\penalty0 (3):\penalty0 355 (2007).

\bibitem[F.~Ginelli(2007)]{Ginelli}
F.~Ginelli, P.~Poggi, A.~Turchi, H.~Chat\'e, R.~Livi, and A.~Politi.
\newblock Characterizing dynamics with covariant Lyapunov vectors.
\newblock \emph{Phys. Rev. Lett.}, 99:\penalty0 130601 (2007).

\bibitem[V.I.~Oseledets(1968)]{osel}
V.I.~Oseledet.
\newblock Multiplicative ergodic theorem: Characteristic Lyapunov exponents of  dynamical systems.
\newblock \emph{Trudy MMO}, 19:\penalty0 179 (1968).

\bibitem[D.~Ruelle(1979)]{ruel}
D.~Ruelle.
\newblock Ergodic theory of differentiable dynamical systems.
\newblock \emph{IHES Publ. Math.}, 50:\penalty0 27 (1979).

\bibitem[S.~Tomsovic(2007)]{Tom}
S.~Tomsovic and A.~Lakshminarayan.
\newblock Fluctuations of finite-time stability exponents in the standard map and the detection of small islands.
\newblock \emph{Phys. Rev. E}, 76:\penalty0 036207 (2007).

\bibitem[C.~Manchein(2009)]{Ces}
C.~Manchein and R.~Artuso.
\newblock Instability statistics and mixing rates.
\newblock \emph{Phys. Rev. E}, 80:\penalty0 036210 (2009).

\bibitem[K.~Bloor(2009)]{L1}
S.~Luzzatto and K.~Bloor.
\newblock Some remarks on the geometry of the standard map.
\newblock \emph{Int. J. Bif. Chaos}, 19:\penalty0 2213, (2009).

\bibitem[M.~Holland(2006)]{L2}
S.~Luzzatto and M.~Holland.
\newblock Stable manifolds under very weak hyperbolicity conditions.
\newblock \emph{Jour. of Diff. Eq.}, 221:\penalty0 444, (2006).

\bibitem[I.~Melbourne(2009)]{Mel}
I.~Melbourne.
\newblock Large and moderate deviations for slowly mixing dynamical systems.
\newblock \emph{Proc.Amer.Math.Soc.}, 137:\penalty0 1735, (2009).

\bibitem[E.M.~McMillan(1967)]{mcmill}
E.M.~McMillan.
\newblock \emph{Some Thoughts on Stability in Nonlinear Periodic Focusing Systems.}
\newblock University of California Radiation Laboratory, Lawrence Berkeley National Laboratory,
U.S. Department of Energy (1967).


\bibitem[]{foot}
\newblock Equation (\ref{stan}) is actually the inverse of McMillan’s original form but, given its time-symmetry, the two are equivalent.

\bibitem[B.~Chirikov(1969)]{Chirikov}
B.~Chirikov.
\newblock Research concerning the theory of nonlinear resonance and stochasticity.
\newblock \emph{Institute of Nuclear Physics, Novosibirsk, Preprint}, \penalty0
  (267), (1969).

\bibitem[A.J.~Lichtenberg(1992)]{lieb}
M.A.~Lieberman and A.J.~Lichtenberg.
\newblock \emph{Regular and Chaotic Dynamics.}
\newblock Applied Mathematical Sciences, Volume 38, Springer (1992).

\bibitem[M.~Sala(2012)]{Proc}
M.~Sala, C.~Manchein and R.~Artuso.
\newblock Estimating hyperbolicity of chaotic bidimensional maps.
\newblock \emph{Int.J.Bif. \& Chaos}, 22, 1250217, (2012).

\bibitem[L.~Barreira(2002)]{Pesin}
Y.B.~Pesin and L.~Barreira.
\newblock \emph{Lyapunov exponents and smooth ergodic theory.}
\newblock University Lecture Series, v. 23, AMS, Providence, (2001).

\bibitem[Sinai(1999)]{Sinai}
Ya.~G. Sinai.
\newblock \emph{A mechanism of ergodicity in standard-like maps. (extended abstract)}.
\newblock Hamiltonian Systems with Three or More Degrees of Freedom,
NATO ASI Series, Volume 533, pp. 242-243, (1999).

\bibitem[Devaney(1976)]{dev}
R.~L. Devaney.
\newblock Reversible diffeomorphisms and flows.
\newblock \emph{Trans. Am. Math. Soc.}, 218:\penalty0 89--113, (1976).

\bibitem[G.~Benettin(1980)]{Benettin}
J.M.~Strelcyn, G.~Benettin and L.~Galgani.
\newblock Lyapunov characteristic exponents for smooth dynamical systems and for Hamiltonian systems; a method for computing all of them.
\newblock \emph{Meccanica}, 15\penalty0 (1):\penalty0 9, (1980).

\bibitem[A.~Adrover(1999)]{Giona}
M.~Giona and A.~Adrover.
\newblock Geometric properties of quasi-periodic orbits of 2d Hamiltonian.
\newblock \emph{Phys. Lett. A}, 259:\penalty0 451, (1999).

\bibitem[J.M.~Steele(1989)]{king}
J.M.~Steele.
\newblock Kingman's subadditive ergodic theorem.
\newblock \emph{AIHP} (B) Probabilités et Statistiques, 25, 1:\penalty0 93--98, (1989).

\bibitem[S.~Cerbelli(2000)]{cerb}
F.J.~Muzzio, S.~Cerbelli and J.M.~Zalc.
\newblock The evolution of material lines curvature in deterministic chaotic flows.
\newblock \emph{Chem. Eng. Sci.}, 55, 2:\penalty0 363--371, (2000).

\bibitem[J.L.~Thiffeault(2001)]{thiff1}
A.H.~Boozer and J.L.~Thiffeault.
\newblock Geometrical constraints on finite-time Lyapunov exponents in two and three dimensions.
\newblock \emph{Chaos}, 11\penalty0 (1):\penalty0 16--28, (2001).

\bibitem[F.~Giovannini(1991)]{Politi}
A.~Politi and F.~Giovannini.
\newblock Homoclinic tangencies, generating partitions and curvature of invariant manifolds.
\newblock \emph{J. Phys. A}, 24:\penalty0 1837, (1991).

\bibitem[J.L.~Thiffeault(2004)]{thiff3}
J.L.~Thiffeault.
\newblock Stretching and curvature of material lines in chaotic flows.
\newblock \emph{Physica D}, 198:\penalty0 169--181, (2004).

\bibitem[F.~Christiansen(1995)]{chris1}
F.~Christiansen and A.~Politi
\newblock A generating partition for the standard map.
\newblock \emph{Phys. Rev. E} 51, 3811 (1995)

\bibitem[F.~Christiansen(1996)]{chris2}
F.~Christiansen and A.~Politi
\newblock Symbolic encoding in symplectic maps.
\newblock \emph{Nonlinearity} 9, 1623 (1996)

\bibitem[P.L.~Garrido(1997)]{garr}
P.L.~Garrido.
\newblock Kolmogorov-Sinai entropy, Lyapunov exponents, and mean free time in billiard systems.
\newblock \emph{J.~Stat.~Phys.},\penalty0 Vol. 88, No. 3--4, pp. 807-824 (1997).

\newpage

\bibitem[N.~Chernov(2001)]{cher}
N.~Chernov and R.~Markarian.
\newblock Introduction to Ergodic Theory of Chaotic Billards.
\newblock \emph{Pub. Mat. Rio de Janeiro}:\penalty0 IMPA, 2001.




\end{thebibliography}

\end{document}